\documentclass[twocol]{ametsoc}

\journal{jcli}

\bibpunct{(}{)}{;}{a}{}{,}

\title{The role of elevated terrain and the Gulf of Mexico in the production of severe local storm environments over North America\\ {\color{red}Revised for Journal of Climate, under review}} 

    \authors{Funing Li\correspondingauthor{Funing Li, 
    Purdue University, Department of Earth, Atmospheric, and Planetary Sciences, 550 Stadium Mall Drive, West Lafayette, IN 47907.}}
    \affiliation{Purdue University, Department of Earth, Atmospheric, and Planetary Sciences, West Lafayette, IN}
    \email{li3226@purdue.edu}
    
    \extraauthor{Daniel R. Chavas}
    \extraaffil{Purdue University, Department of Earth, Atmospheric, and Planetary Sciences, West Lafayette, IN}

    \extraauthor{Kevin A. Reed}
    \extraaffil{School of Marine and Atmospheric Sciences, Stony Brook University, Stony Brook, NY}
    
    \extraauthor{ Nan Rosenbloom}
    \extraaffil{National Center for Atmospheric Research, Boulder, CO}
    
    \extraauthor{ Daniel T. Dawson II}
    \extraaffil{Purdue University, Department of Earth, Atmospheric, and Planetary Sciences, West Lafayette, IN}
	
\abstract{The prevailing conceptual model for the production of severe local storm (SLS) environments over North America asserts that upstream elevated terrain and the Gulf of Mexico are both essential to their formation. This work tests this hypothesis using two prescribed-ocean climate model experiments with North American topography removed or the Gulf of Mexico converted to land and analyzes how SLS environments and associated synoptic-scale drivers (southerly Great Plains low-level jets, drylines, elevated mixed layers, and extratropical cyclones) change relative to a control historical run. Overall, SLS environments depend strongly on upstream elevated terrain but weakly on the Gulf of Mexico. Removing elevated terrain substantially reduces SLS environments especially over the continental interior due to broad reductions in both thermodynamic and kinematic parameters, leaving a more zonally-uniform residual distribution that is maximized near the Gulf coast and decays toward the continental interior. This response is associated with a strong reduction in synoptic-scale drivers and a cooler and drier mean-state atmosphere. Replacing the Gulf of Mexico with land modestly reduces SLS environments thermodynamically over the Great Plains and increases them kinematically over the eastern U.S, shifting the primary local maximum eastward into Illinois; it also eliminates the secondary, smaller local maximum over southern Texas. This response is associated with modest changes in synoptic-scale drivers and a warmer and drier lower-tropospheric mean state. These experiments provide insight into the role of elevated terrain and the Gulf of Mexico in modifying the spatial distribution and seasonality of SLS environments.}

\begin{document}

\newcommand{\todo}[1]{\textcolor{red}{\textbf{[#1]}}}

\maketitle


\section{Introduction}\label{sec:introduction}

North America is perhaps the most prominent hot-spot globally for severe local storm (SLS) events, including those that produce damaging winds, large hailstones, and/or tornadoes \citep{ludlam1963, johns1992}. Though SLS events are small scale, they develop principally within favorable larger-scale environments. These environments are commonly defined using proxies that combine key thermodynamic and kinematic ingredients, particularly: 1) the product of convective available potential energy (CAPE) and 0--6-km bulk vertical wind shear (S06), and 2) the 0--3-km energy helicity index (EHI03) that is proportional to the product of CAPE and 0--3-km storm relative helicity (SRH03) \citep{Rasmussen_Blanchard_1998, Rasmussen_2003, Brooks_etal_2003, Doswell_Schultz_2006, Grams_etal_2012}. To trigger deep convection, air parcels must first overcome a stable layer defined by high convective inhibition (CIN), which suppresses upward motion and hence allows CAPE to build up until convection is initiated \citep{colby1984,williams1993,Agard_Emanuel_2017,chen2020}. Large-scale mean ascent and synoptic-scale dynamics such as convergence along air mass boundaries (e.g., synoptic fronts and drylines) and orographic lifting are efficient ways to initiate convective storms \citep{markowski2011}. SLS environments over North America are generally confined to the eastern half of the U.S., especially the Great Plains \citep{Brooks_etal_2003,gensini_ashley_2011,li2020}, though recent studies have indicated a tendency of eastward shift of these environments in past decades \citep{Gensini_Brooks_2018, tang2019}. 

The climatology of SLS environments, including strong seasonal and diurnal cycles, generally perform well in capturing statistical variability in SLS activity itself in the U.S. \citep{gensini_ashley_2011, Agee_etal_2016, Gensini_Brooks_2018, tang2019}. Thus, improving our understanding of what generates SLS environments in the first place can help us better understand SLS activity on climate time-scales. Moreover, the larger-scale nature of these environments allows for the use of reanalysis data \citep{Brooks_etal_2003,gensini_ashley_2011,allen2014, Gensini_etal_2014, Tippett_etal_2016, Gensini_Brooks_2018,Taszarek_etal_2018,tang2019,taszarek2020_01,taszarek2020_02,taszarek2020_03} and global climate models \citep{Diffenbaugh_etal_2013,tippett_2015,chen2020,li2020}, which can resolve SLS environments though not actual SLS events. To date, though, global climate model experiments have yet to be applied to test how SLS environments are generated over North America in the present climate state in the first place, which may limit our ability to predict how SLS activity may change due to climate change.

\begin{figure}[t]
\centerline{\includegraphics[width=19pc]{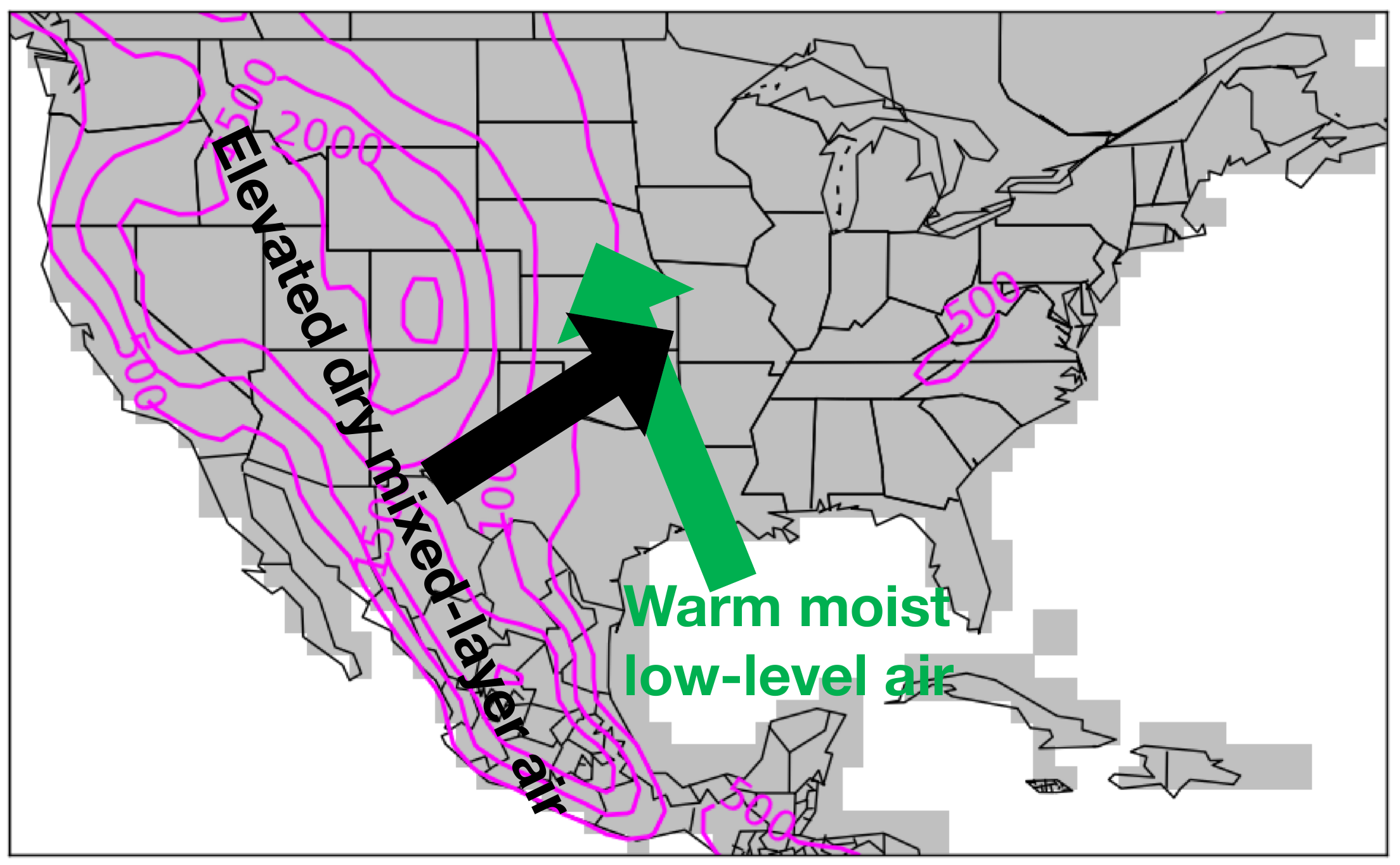}}
\caption{Schematic diagram of the conceptual model by \citet{Carlson_etal_1983} representing conditions favorable for the formation of severe local storm environments over the central U.S. Contour lines indicate elevation (m).}
\label{fig_model}
\end{figure}

The prevailing conceptual model for the generation of SLS environments over the eastern U.S. was proposed by \citet{Carlson_etal_1983} (Figure \ref{fig_model}). This model identifies elevated terrain to the west and the Gulf of Mexico to the south as the key geographic features essential to producing these environments. First, the Gulf of Mexico provides a source of warm and moist low-level air. Second, surface heating of the Colorado and Mexican plateaus upstream generates dry adiabatic layers aloft characterized by steep lapse rates. SLS environments then arise downstream of the elevated terrain due to the superposition of these two layers. In reality, this superposition is typically mediated by synoptic-scale features including southerly Great Plains low-level jets (GPLLJs), drylines, elevated mixed layers (EMLs), and extratropical cyclones. Specifically, southerly GPLLJs enhance the inland transport of warm and moist low-level air from the Gulf of Mexico, especially during the nighttime in spring and summer \citep{Bonner_1968,Helfand_Schubert_1995, Whiteman_etal_1997,Higgins_etal_1997,weaver2008, weaver2012}. Meanwhile, the westerly jet stream advects the well-mixed air aloft downstream of the elevated terrain to form meridionally oriented drylines over the Great Plains when the descending dry air encounters the moist near-surface air from the Gulf of Mexico \citep{Fujita_1958, Schaefer_1974, Ziegler_Hane_1993, Hoch_Markowski_2005}. The canonical synoptic flow patterns associated with severe weather outbreaks over the Great Plains also highlights the roles of a southwesterly jet aloft and the southerly GPLLJ extending inland from the Gulf of Mexico downstream of the elevated terrain \citep{barnes1986, johns1992, johns1993, Mercer_etal_2012, li2020}. Farther east, advection of the well-mixed layer from the elevated terrain produces an EML over the moist low-level air that commonly forms strong capping inversions and creates CIN that inhibits convective initiation for boundary layer parcels \citep{Carlson_etal_1983,Lanicci_Warner_1991a,Banacos_Ekster_2010}. This can allow for a strong buildup of CAPE during the daytime heating \citep{Carlson_etal_1983,Farrell_Carlson_1989, Lanicci_Warner_1991b,Lanicci_Warner_1991c, Cordeira_etal_2017,Ribeiro_Bosart_2018}. Moreover, the above model implicitly assumes differential advection of two layers and thus is associated with vertical wind shear. These CAPE- and shear-producing processes are often strongly amplified locally in the presence of a surface extratropical cyclone \citep{Doswell_Bosart_2001,Hamill_etal_2005, Tochimoto_etal_2015}, whose formation is favored downstream of the Rocky Mountains \citep{held2002, brayshaw2009}. In combination, the result is the production of SLS environments characterized by high values of CAPE and shear.

Previous studies partially tested this conceptual model using limited-area numerical model experiments but focused principally on the role of elevated terrain \citep{benjamin1986, benjamin1986b, arritt1992, rasmussen2016}. \citet{benjamin1986} tested the key role of elevated terrain for two historical SLS outbreaks over the Great Plains in a regional mesoscale modeling framework. \citet{benjamin1986b} and \citet{arritt1992} further examined effects of elevated terrain on the production of strong capping inversions downstream based on two-dimentional idealized experiments. \citet{rasmussen2016} found a strong orographic control on convective initiation downstream of the Andes over South America via high-resolution simulations of a convective system from the Weather Research and Forecasting Model, and indicated a conceptual model for the production of SLS environments in subtropical South America that is similar to the U.S. Great Plains. Additionally, various numerical model experiments with terrain modifications have shown substantial impacts of orography on larger-scale atmospheric flows including GPLLJs \citep{pan2004, ting2006} and stationary waves or storm tracks \citep{broccoli1992, held2002, inatsu2002, brayshaw2009, chang2009, wilson2009, sandu2019,lutsko2019}. However, this conceptual model, including both upstream elevated terrain and the Gulf of Mexico, has yet to be tested using global climate model experiments.

Thus, this work aims to explicitly test the conceptual model of \citet{Carlson_etal_1983} using a global climate model (Community Atmosphere Model version 6, CAM6) by addressing the following questions:

\begin{enumerate}
\item{Are western U.S. elevated terrain and the Gulf of Mexico each necessary conditions for producing SLS environments over North America? Do they affect the seasonality of these environments?}
\item {How are these responses of SLS environments associated with responses of SLS-relevant synoptic-scale features in each experiment?}
\item {How are these responses of SLS environments and SLS-relevant synoptic-scale features associated with changes in the mean state and characteristic synoptic flow patterns in each experiment?}
\end{enumerate}

To answer these questions, we perform numerical experiments using CAM6, in which we eliminate any topography over North America or convert the Gulf of Mexico to land, respectively. This work serves as a direct assessment of the geographic controls of SLS environments to help better understand the formation of these environments within the Earth's climate system. These experiments are compared against a control historical simulation presented in \citep[][hereafter L20]{li2020}. L20 found that CAM6 can broadly reproduce the climatology of SLS environments and associated synoptic-scale features over North America, as compared to the ERA5 reanalysis dataset. L20 also noted a few key biases in CAM6, including a high bias in CAPE over the eastern third of the U.S. associated with the systematic warm and moist biases that are known to persist across many regional and global climate models \citep{klein2006, cheruy2014, mueller2014, lin2017, Seeley_Romps_2015}.   

Section \ref{sec:methdology} describes our experimental design and analysis methodology. Section \ref{sec:results-notopo} analyzes the responses of SLS environments, the associated synoptic-scale features, and the synoptic flow patterns to removing elevated terrain. Section \ref{sec:results-nogom} analyzes these responses to filling the Gulf of Mexico. Finally, Section \ref{sec:conclusions} summarizes key conclusions and discusses avenues for future work.

\section{Methodology}\label{sec:methdology}

\subsection{Experimental Design}

We use CAM6 as our experimental laboratory for this study. CAM6 is the atmospheric component of the Community Earth System Model version 2.1 (available at \url{ http://www.cesm.ucar.edu/models/cesm2/}) developed in part for participation in the Coupled Model Intercomparison Project 6 \citep{eyring2016}. Compared with its predecessor CAM5 \citep{Neale_etal_2012}, CAM6 contains substantial modifications to the physical parameterization suite: schemes for boundary layer turbulence, shallow convection and cloud macrophysics in CAM5 are replaced by the Cloud Layers Unified by Binormals \cite[CLUBB;][]{golaz2002,bogenschutz2013} scheme; the improved two-moment prognostic cloud microphysics from \citet{gettelman2015}, which carries prognostic precipitation species (rain and snow) in addition to cloud condensates; and the orographic drag parameterizations are also updated. CAM6 uses the \citet{zhang1995} scheme for deep convection parameterization, in which deep convection is triggered when CAPE exceeds 70 J kg$^{-1}$ and a CAPE-based closure assumption is used. In addition, two main modifications are added to the standard \citet{zhang1995} scheme in CAM6. The first enhancement is to include the effect of lateral entrainment dilution into the calculation of CAPE \citep{neale2008}, which allows mixing of the rising plume with surrounding environmental air; the second enhancement is the addition of convective momentum transport \citep{richter2008}. In addition, the \citet{zhang1995} scheme in CAM6 has been retuned compared to earlier versions in CAM predecessors to increase the sensitivity to convective initiation.   

\begin{figure*}[ht]
\centerline{\includegraphics[width=38pc]{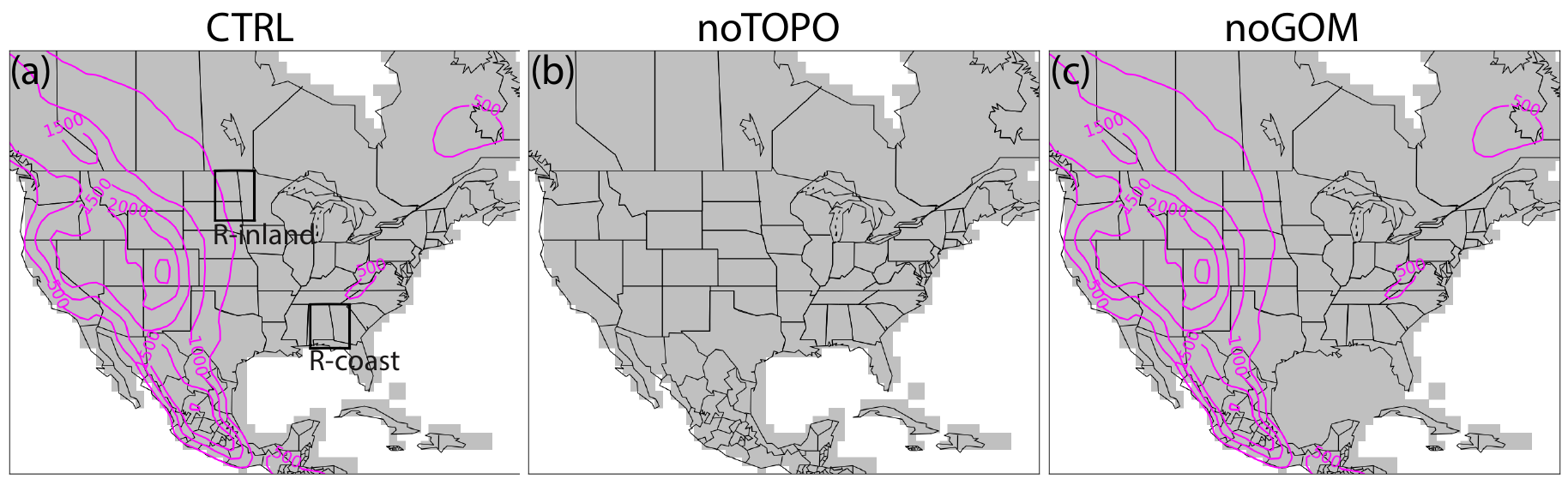}}
\caption{Elevation (m; contour lines) and land mask (filled gray) for (a) CTRL, (b) noTOPO, and (c) noGOM. R-inland and R-coast in (a) denote the two 5$^\circ\times$5$^\circ$ sub-region boxes selected for regional analysis.}
\label{fig_map}
\end{figure*}

We define as our control simulation (CTRL) an Earth-like climate state over the period 1979--2014, following the Atmospheric Model Intercomparison protocols \cite[AMIP;][]{Gates_etal_1999}. This CTRL run is exactly the same simulation we performed and evaluated in L20, in which our analysis has indicated that this CAM6 CTRL broadly reproduces the climatology of SLS environments and the associated synoptic-scale features over North America; this similar set-up was also examined in previous work with CAM5 \citep{wehner2014, varuolo2019}. CAM6 is configured with the default finite volume dynamical core on a 0.9$^\circ\times$1.25$^\circ$ latitude-longitude grid mesh. The simulation reaches its equilibrium state quickly in terms of the steady temporal evolution of surface heat fluxes over the eastern U.S., western U.S., and the Gulf of Mexico, respectively (Figure S1a,b). Thus, we only discard the first year for spinup and analyze the 3-hourly output from 1980--2014.

From CTRL, we perform two experiments to investigate the role of elevated terrain and the Gulf of Mexico in producing SLS environments over North America. In the first experiment, we set North American topography to zero (i.e., elevation$=$0 m, Figure \ref{fig_map}b) without changing land cover type (denoted as ``noTOPO'' experiment). In the second experiment, we convert the Gulf of Mexico to land by replacing ocean cells in the Gulf of Mexico with land (Figure \ref{fig_map}c). In addition to the modified land mask, we set the plant functional type (PFT) to C4 grass in the Community Land Model version 5 \cite[CLM5;][]{lawrence2018} over this ``new land'' for simplicity. This set up can be defined as a low-lying plain covered by grass (denoted as ``noGOM'' experiment). This added grassland strongly increases surface sensible heat flux (from $\sim$10 W m$^{-2}$ in CTRL to up to 100 W m$^{-2}$ in noGOM) and decreases surface latent heat flux (from $\sim$200 W m$^{-2}$ in CTRL to less than 100 W m$^{-2}$ in noGOM) over the Gulf of Mexico during warm seasons (Figure S1a,b vs. e,f), consistent with differences in surface energy budget over land and ocean. Ultimately we do not know what the true land type would be if the Gulf of Mexico were land, and existing research suggests significant intrinsic uncertainty given that subtropical land surfaces can have multiple stable states \citep{rietkerk2011, staver2011}. Future work could test the effects of filling the Gulf of Mexico with different, or a combination of, land types. Similar to CTRL, we perform both noTOPO and noGOM experiments over the period 1979--2014 and analyze the 3-hourly output from 1980--2014 when the model is in steady state (Figure S1c--f).

\begin{table*}[t]
\caption{Identifying criteria for the SLS-relevant synoptic-scale features: southerly GPLLJs, drylines, EMLs, and extratropical cyclone activity including cyclone track and 850-hPa EKE. The reader is referred to \citet{li2020} for detailed explanations.}
\begin{center}
\begin{tabular}{ccccrrcrc}
\hline\hline
$SYNOPTIC$-$SCALE\;\; FEATURES$ & $IDENTIFYING\;\; CRITERIA$ \\
\hline
 Southerly GPLLJ & (1) Maximum wind speed below 3000 m: $V_{max}\geq$ 10 m s$^{-1}$ \\
                 & (2) Largest decrease from $V_{max}$ to wind at 3000 m: $\Delta V\geq$ 5 m s$^{-1}$\\
                 & (3) Direction of $V_{max}$ falls between 113$^\circ$--247$^\circ$ \\
 Dryline & (1) Horizontal gradient of surface specific humidity: ${\nabla}_{H}{q}\geq 3\times10^{-5}$ km$^{-1}$ and $\frac{\partial{q}}{\partial{x}}>$0 \\
         & (2) Surface temperature: $\frac{\partial{T}}{\partial{x}}<$0.02 K km$^{-1}$ \\
         & (3) Surface wind direction on the west side is 170$^\circ$--280$^\circ$ and on the east side 80$^\circ$--190$^\circ$ \\
 EML & (1) A layer with lapse rate $\geq$ 8.0 K km$^{-1}$ through a depth of at least 200 hPa \\ 
     & (2) Relative humidity increases from the base to the top of the layer \\ 
     & (3) The base is higher than 1000 m but below 500-hPa level \\ 
     & (4) Average lapse rate between the base and surface $\leq$8.0 K km$^{-1}$ \\
 Cyclone track & (1) Candidate cyclones are minima in SLP with a closed contour 2 hPa greater than the minimum \\
             & (2) The closed contour lies within 6 great circle degrees of the minimum \\
             & (3) Candidate cyclones are then stitched together in time by searching within an 8-degree great circle \\
             & radius at the next time increment for another candidate cyclone to form a cyclone track \\
             & (4) A cyclone track must exist for at least 24 hours \\
  850-hPa EKE  & (1) Determine $(u\;',\; v\;')$--zonal and meridional velocity deviation from annually mean velocities \\
               & (2) Apply a 2--6-day Butterworth bandpass filter to $(u\;',\; v\;')$ \\
               & (3) {EKE = 0.5 $(u\;'^2+v\;'^2)$} \\
\hline
\end{tabular}
\end{center}
\end{table*}      

\subsection{Analysis}

\subsubsection{SLS Environments}

Our analyses focus on responses of the climatology of SLS environments in noTOPO and noGOM each as compared to CTRL. We define SLS environments as the 99th percentile of two combined proxies, CAPES06 \citep{Brooks_etal_2003} and EHI03 \citep{Hart_Korotky_1991, Davies_1993}, as well as their constituent parameters. Specifically, CAPES06 is the product of surface-based CAPE \citep{Doswell_Rasmussen_1994} and S06 \citep{Rasmussen_Blanchard_1998, Weisman_Rotunno_2000}; EHI03 is 
a dimensionless quantity proportional to the product of surface-based CAPE and SRH03 \citep{Davies_1990}. We also analyze surface-based CIN \citep{colby1984,williams1993,riemann2009}, which permits the build up of CAPE. Calculations of these proxies and parameters use the following equations: 
\begin{equation} {CAPES06 = CAPE\cdot S06}, \end{equation} 
\begin{equation} {EHI03 = \frac{CAPE\cdot SRH03}{160,000\ m^{4}\ s^{-4}}}, \end{equation}
\begin{equation} CAPE={\int_{z_{LFC}}^{z_{EL}} g\frac{T_{vp}-T_{ve}}{T_{ve}} dz}, \end{equation}
\begin{equation} S06={|\textbf{V}_{6km}-\textbf{V}_{10m}|}, \end{equation}
\begin{equation} SRH03={-\int_{z_b}^{z_t} \hat{\textbf{k}}\cdot(\textbf{V}-\textbf{C})\times\frac{\partial\textbf{V}}{\partial{z}} dz}, \end{equation}
\begin{equation} CIN={-\int_{z_{p}}^{z_{LFC}} g\frac{T_{vp}-T_{ve}}{T_{ve}} dz}, \end{equation}
where $g=$ 9.81 m s$^{-2}$ is the acceleration due to gravity, $z_{LFC}$ and $z_{EL}$ denote the level of free convection and the equilibrium level, $z_{p}=2$ m indicates the 2-m parcels, $T_{vp}$ and $T_{ve}$ are virtual temperature of the 2-m parcels and the environment as a function of height, $\textbf{V}_{6km}$ and $\textbf{V}_{10m}$ are horizontal wind vectors at 6 km and 10 m AGL, \textbf{V} is the horizontal wind vector as a function of height, \textbf{C} is the storm motion vector following the definition and calculation from \citet{bunkers_2000}, $z_b=10$ m is the altitude of the layer bottom, $z_t=3$ km is the altitude of the layer top, and $\hat{\textbf{k}}$ is the vertical unit vector. 

To generate climatological distributions of these SLS environmental proxies and parameters over North America, we first use Eqs.(1)--(6) to calculate each for the period 1980--2014 from 3-hourly model outputs. We then calculate their 99th percentiles for each grid point based on their time series during March--August, as SLS environments in general peak in spring and summer \citep{Diffenbaugh_etal_2013, li2020}. Considering that the specific seasonal phase (peak month) of these environments varies regionally, we also analyze changes in their seasonal cycle using their monthly 99th percentiles for each experiment. Though L20 noted an overestimation of SLS environments over the eastern U.S. in CTRL, this may not affect their responses in each experiment, as these biases are associated with systematic model biases of temperature and moisture in CAM6 and thus may persist in the experiments.   

We also seek to understand how changes in the constituent parameters ($\Delta{CAPE}$, $\Delta{S06}$, or $\Delta{SRH03}$) affect changes in the combined proxies ($\Delta{CAPES06}$ or $\Delta{EHI03}$). We decompose fractional changes in $CAPES06$ according to 
\begin{equation} 
\begin{split}
\frac{\Delta{CAPES06}}{CAPES06_{CTRL}} & = \frac{\Delta{CAPE}}{CAPE_{CTRL}}+\frac{\Delta{S06}}{S06_{CTRL}} \\
     & +\frac{\Delta{CAPE}}{CAPE_{CTRL}}\cdot \frac{\Delta{S06}}{S06_{CTRL}}
\end{split}
\end{equation}
where $CAPE_{CTRL}$ and $S06_{CTRL}$ are $CAPE$ and $S06$ associated with $CAPES06$ in CTRL ($CAPES06_{CTRL}$). $\Delta$ denotes the difference between each experiment and CTRL (i.e., noTOPO or noGOM minus CTRL). The term on the left hand side of Eq. (7) is the fractional change of $CAPES06$ in each experiment relative to CTRL. The first and second terms on the right hand side represent the fractional changes due to changes in the associated $CAPE$ and $S06$, and the third term is a second-order residual term that is generally small. This decomposition method is similar to a statistical framework developed to isolate the dynamic and thermodynamic components of cloud changes \citep{Bony_etal_2004} or extreme precipitation \citep{Emori_Bony_2005, chen&chavas2020}. Similarly, the fractional change in $EHI03$ is given by
\begin{equation} 
\begin{split}
\frac{\Delta{EHI03}}{EHI03_{CTRL}} & = 
\frac{\Delta{CAPE}}{CAPE_{CTRL}}+\frac{\Delta{SRH03}}{SRH03_{CTRL}} \\
     & +\frac{\Delta{CAPE}}{CAPE_{CTRL}}\cdot \frac{\Delta{SRH03}}{SRH03_{CTRL}}
\end{split}
\end{equation}

\subsubsection{SLS-relevant synoptic-scale features}

To understand responses of SLS environments, we then analyze changes in key synoptic-scale features associated with the formation of SLS environments over North America, including southerly GPLLJs, drylines, EMLs, and extratropical cyclone activity. We follow L20 and references therein to identify GPLLJs \citep{Bonner_1968,Whiteman_etal_1997,Walters_etal_2008, Doubler_etal_2015}, drylines \citep{Hoch_Markowski_2005,Duell_Broeke_2016}, EMLs \citep{Banacos_Ekster_2010, Ribeiro_Bosart_2018}, and extratropical cyclone activity including cyclone track \citep{Ullrich_Zarzycki_2017, zarzycki2018} and the 2--6-day Butterworth bandpass filtered eddy kinetic energy (EKE) at 850 hPa \citep{blackmon_1976, russell_2006, Ulbrich_2008, Harvey_2014, Schemm_2018}. Identifying criteria for each are summarized in Table 1; the reader is referred to L20 for detailed explanations. We analyze the responses of mean occurrence frequency of these synoptic-scale features during March--August of 1980--2014 in noTOPO and noGOM relative to CTRL.

\subsubsection{Mean State and Characteristic Synoptic Flow Pattern}

To better understand responses of SLS environments and the associated synoptic-scale features, we analyze differences in the mean-state atmosphere for each experiment. We examine changes of low-level moisture and temperature (925 and 850 hPa), as they are crucial to CAPE response; we also examine changes of the horizontal flow field at lower and upper levels (925, 850, and 250 hPa) relevant to the S06 or SRH03 response \citep{Trapp_etal_2007, Diffenbaugh_etal_2013,Agard_Emanuel_2017,chen2020,li2020}. In addition, responses of the mean 925-hPa moisture transport and 850-hPa wind and temperature, especially during warm seasons, may be associated with changes in GPLLJs and EMLs \citep{pan2004, ting2006, Ribeiro_Bosart_2018}. Meanwhile, responses in the mean 250-hPa wind speed indicate shifts in the jet stream relevant to extratropical cyclone activity \citep{holton_1973}.  

Additionally, we evaluate changes in the composite synoptic patterns associated with extreme SLS environments to examine the extent to which mean-state responses are also found in the characteristic synoptic flow responses associated with extreme SLS environments. As past work has indicated classic synoptic patterns of SLS events and environments over much of the eastern U.S. \citep{barnes1986,johns1992,johns1993,Mercer_etal_2012,li2020}, our analysis also serves to examine if these common synoptic patterns are sensitive to the existence of elevated terrain and the Gulf of Mexico. Here we focus on two sub-regions, R-inland and R-coast, defined in Figure \ref{fig_map}a. Region R-inland represents an inland region over the northern Great Plains where the SLS environments reach a local maximum in CTRL. The associated composite synoptic patterns over R-inland in CTRL are similar to the classic synoptic patterns of severe weather events \citep{li2020}. Region R-coast represents a coastal region over the southeastern U.S. where responses of SLS environments in experiments are relatively small (detailed below in Results). The associated characteristic synoptic patterns over R-coast in CTRL are known to be different from the classic synoptic patterns over the Great Plains \citep{li2020}. Following L20, we identify an extreme case in a region during March--August when the CAPES06 exceeds its local 99th percentile in at least 80\% of the total grid points within the region. Composite synoptic patterns are generated by averaging variable fields from the identified cases. Specifically, we analyze the composite patterns of horizontal wind speed and geopotential height at 250 hPa, temperature and geopotential height at 850 hPa, and near surface properties including 925-hPa specific humidity and wind, and sea level pressure.   
 
\section{Results: Responses to removing North American topography (noTOPO)}\label{sec:results-notopo}

We first analyze responses of SLS environments to the removal of elevated terrain by comparing noTOPO with CTRL. To better understand why these responses occur, we also examine differences in synoptic-scale features (southerly GPLLJs, drylines, EMLs, and extratropical cyclone activity) that frequently generate these environments, as well as differences in the mean state and characteristic synoptic patterns.

\subsection{SLS Environments} 

\begin{figure}[t]
\centerline{\includegraphics[width=19pc]{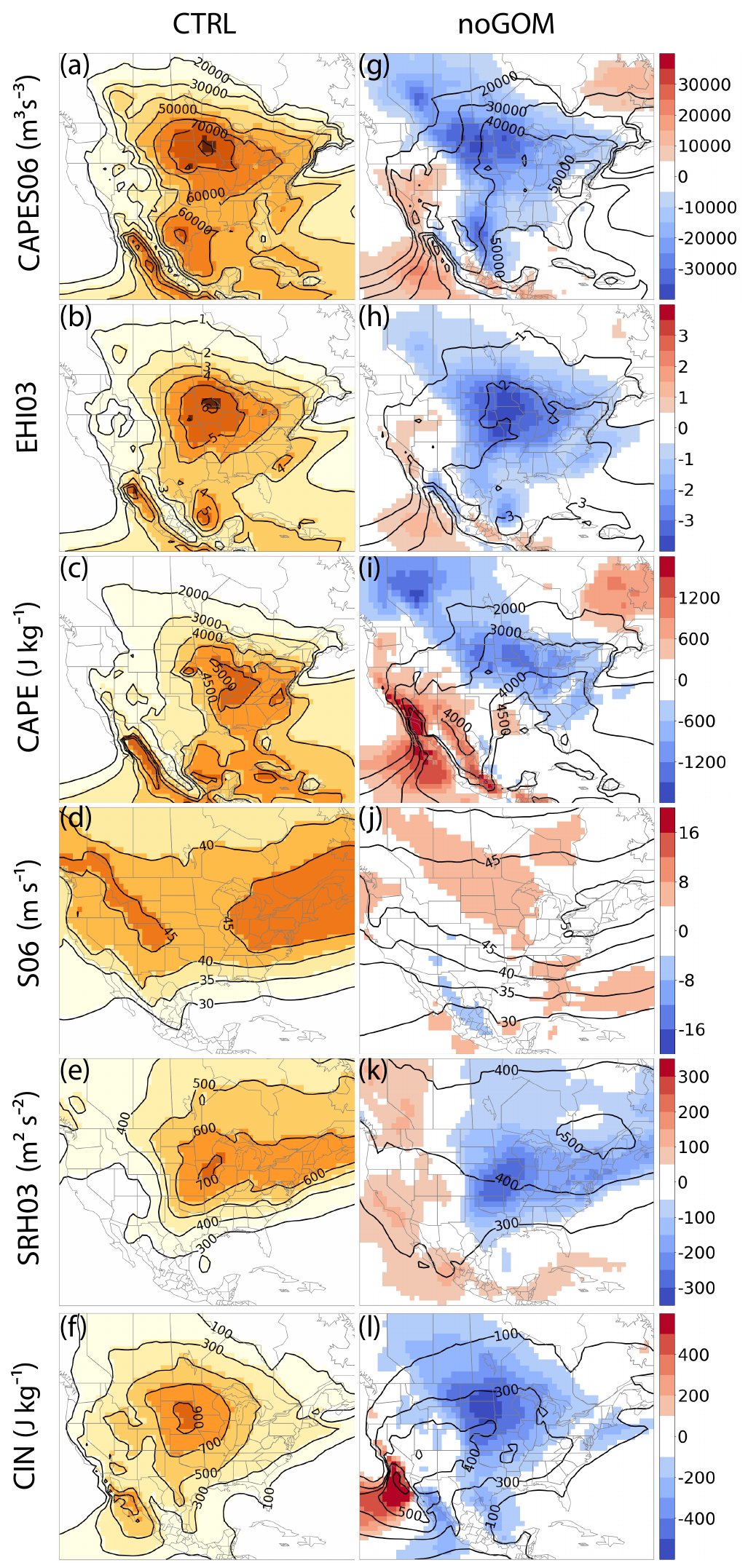}}
\caption{(a--f) CTRL 99th percentiles (contour lines + filled contours) of CAPES06, EHI03, CAPE, S06, SRH03, and CIN; (g--l) as in (a--f) but for noTOPO 99th percentiles (contour lines) and responses (noTOPO minus CTRL; filled contours). The 99th percentiles are generated at each grid point from the 3-hourly full-period (1980--2014) cases during March--August.}
\label{fig_99ile_annual_notopo}
\end{figure}

We begin with analyzing responses of extreme values (99th percentiles) of SLS environmental proxies and parameters during March--August over North America in noTOPO (Figure \ref{fig_99ile_annual_notopo}g--l), as compared to CTRL (Figure \ref{fig_99ile_annual_notopo}a--f). Removing elevated terrain strongly reduces CAPES06 and EHI03 over much of the eastern half of the U.S., with the reduction extending into south-central and southwest Canada (Figure \ref{fig_99ile_annual_notopo}g,h). The largest decrease occurs near the local maxima in CTRL over the northern Great Plains and southern Texas for CAPES06 ($\sim-$30,000 m$^{3}$ s$^{-3}$) and in the central Great Plains for EHI03 ($\sim-$4). The reduction along the east coast and the southeastern U.S., which may be partly due to the removal of the Appalachian Mountains, is relatively small. CAPE is broadly reduced ($\sim-1000$ J kg$^-1$) over a swath stretching from southwestern Canada across the Great Plains to the southeastern U.S., consistent with decreases in CAPES06 and EHI03, whereas it remains similar to CTRL or is slightly enhanced over the south-central U.S. (Figure \ref{fig_99ile_annual_notopo}i). Removing elevated terrain causes S06 and SRH03 to become substantially more zonally symmetric, especially over the Great Plains (Figure \ref{fig_99ile_annual_notopo}j,k). Specifically, S06 slightly increases ($\sim+$6 m s$^{-1}$) along the United States-Canada border east of the Rocky mountains. SRH03 decreases over the eastern half of the U.S. with strongest reductions ($\sim-200$ m$^{2}$ s$^{-2}$) over the central Great Plains. Finally, removing elevated terrain strongly reduces CIN over the Great Plains with the peak reduction over the northern Great Plains (Figure \ref{fig_99ile_annual_notopo}l); the response pattern is similar to CAPES06 and EHI03 and broadly consistent with reductions in CAPE.

\begin{figure*}[t]
\centerline{\includegraphics[width=36pc]{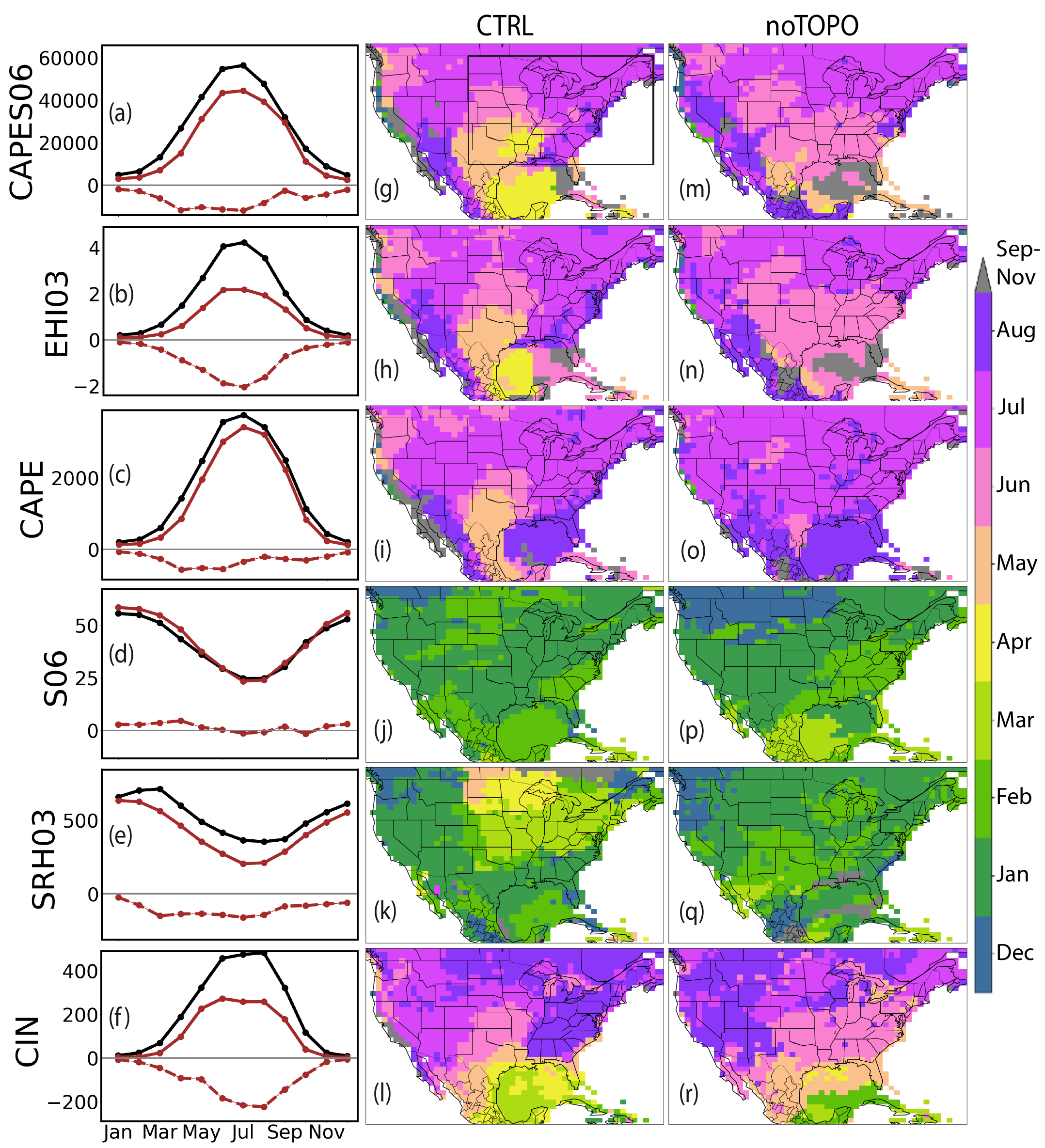}}
\caption{(a--f) The monthly 99th percentiles of CAPES06, EHI03, CAPE, S06, SRH03, and CIN for CTRL (black solid line) and noTOPO (red solid line), and differences in noTOPO relative to CTRL (red dashed line), averaged for land grid points over the eastern half of the U.S. (within the black box in (g)). (g--l) Peak month with the maximum monthly 99th percentiles of each parameter for CTRL; (m--r) as in (g--l) but for noTOPO.}
\label{fig_99ile_peakmm_notopo}
\end{figure*}

SLS environments exhibit a strong seasonal cycle, and the responses of extreme CAPES06 and EHI03 occur principally in the warm seasons and hence reduce their seasonal cycle magnitudes (Figure \ref{fig_99ile_peakmm_notopo}a,b). The seasonal response of CAPES06 is predominantly tied to CAPE, as the response of S06 is consistently small throughout the year with a slight decrease in summer and a slight increase in winter and spring (Figure \ref{fig_99ile_peakmm_notopo}c,d). The seasonal response of EHI03 is tied to both CAPE and SRH03, as both are strongly reduced in the warm seasons (Figure \ref{fig_99ile_peakmm_notopo}b,c,e). A similar dampening of the seasonal cycle is also found in the response of CIN (Figure \ref{fig_99ile_peakmm_notopo}f).

In addition to the magnitude of seasonal cycle, the spatial distribution of the seasonal phase (peak month) of these environments is also modified (Figure \ref{fig_99ile_peakmm_notopo}g--r). In CTRL, the seasonal cycles of CAPES06 and EHI03 both peak in April/May near the Gulf Coast and then progresses inland towards the continental interior through July; this progression is confined zonally to the region downstream of the Rockies over the Great Plains.  Removing elevated terrain causes CAPES06 and EHI03 to peak in June over the entire southern Great Plains and southeastern U.S. (Figure \ref{fig_99ile_peakmm_notopo}m,n). This peak is 1--2 months later than CTRL over the southern Great Plains while 1--2 months earlier over the southeastern U.S. (Figure \ref{fig_99ile_peakmm_notopo}m,n vs. g,h). Hence, elevated terrain appears responsible for the zonal variability in the seasonality of SLS environments over the southeastern U.S. in the real world. This response is tied to CAPE, which exhibits a similar pattern in its response particularly over the southern Great Plains, with peak month shifting from late spring (May--June in CTRL; Figure \ref{fig_99ile_peakmm_notopo}i) to late summer (July--August in noTOPO; Figure \ref{fig_99ile_peakmm_notopo}o); the seasonal peak in CAPE becomes zonally-symmetric, occurring in July north of $\sim$31$^o$ N and August to the south. In contrast, the peak month of S06 changes only minimally, occurring in winter (December--February; Figure \ref{fig_99ile_peakmm_notopo}p) similar to CTRL (Figure \ref{fig_99ile_peakmm_notopo}j). Notably, the zonal asymmetry of S06 persists, highlighting how surface thermodynamic variability (land-ocean and SST variability) still generates a stationary wave pattern \citep{kaspi2013}. SRH03 is phase shifted strongly, from spring (March--May in CTRL; Figure \ref{fig_99ile_peakmm_notopo}k) to winter (January--February in noTOPO; Figure \ref{fig_99ile_peakmm_notopo}q) over much of the eastern half of North America, especially over the northern Great Plains and southern Canada. This suggests the weakened influence of the reduced GPLLJs in spring in noTOPO (detailed below). As a result, the seasonal phase of SRH03 becomes more similar to that of S06 driven predominantly by the jet stream. Finally, the peak month of CIN over the eastern third of the U.S. shifts from late summer (August in CTRL; Figure \ref{fig_99ile_peakmm_notopo}l) to early summer (June in noTOPO; Figure \ref{fig_99ile_peakmm_notopo}r), rendering the seasonal peak in CIN more zonally-symmetric east of the Rocky Mountains, similar to CAPE, while still retaining significant meridional variability. Overall, removing elevated terrain suppresses the inland progression of the seasonal cycle of SLS environments and leads to relatively uniform seasonal phase over the eastern U.S.

\begin{figure}[t]
\centerline{\includegraphics[width=19pc]{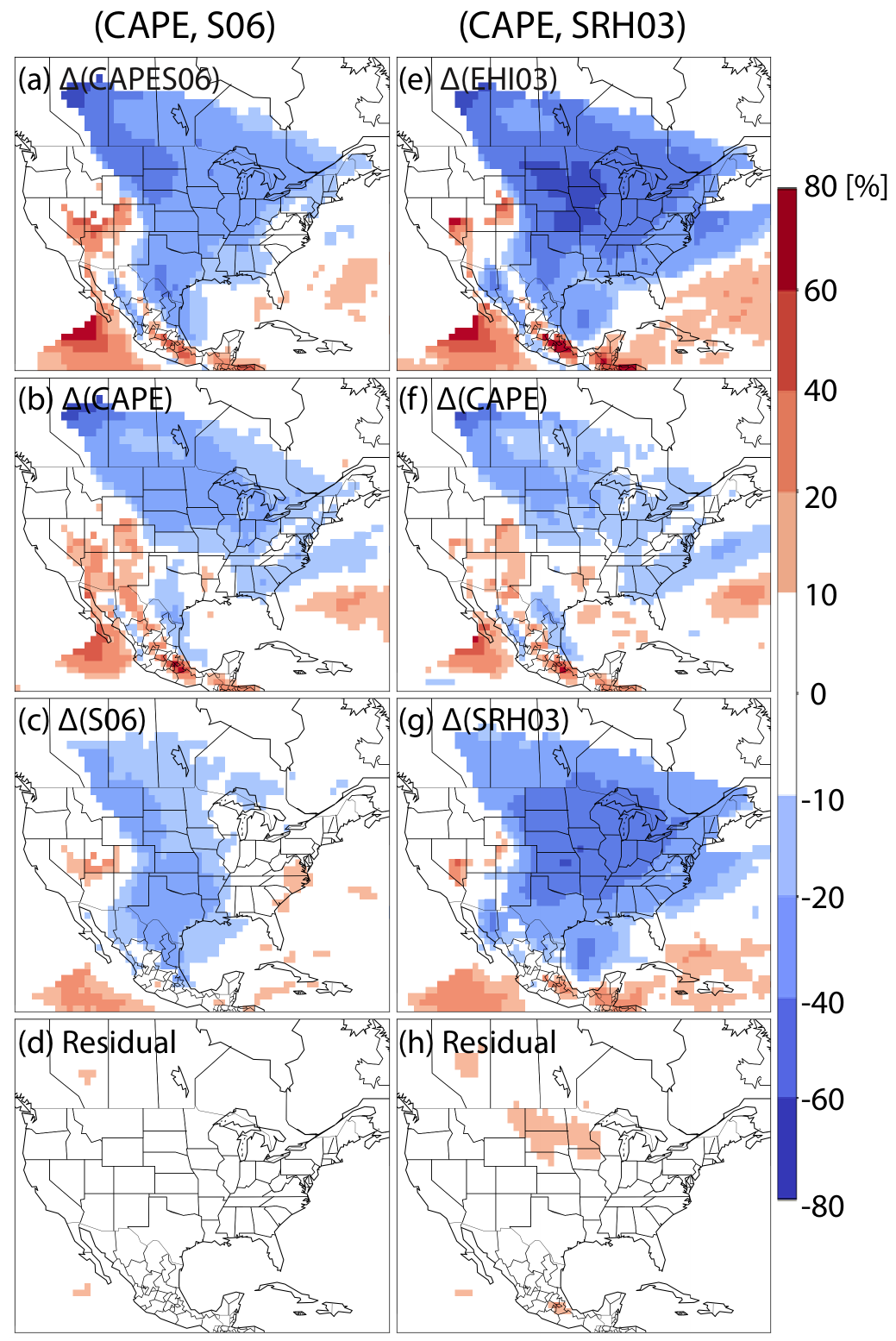}}
\caption{The noTOPO percentage changes (as compared to CTRL) in (a) the center of mass of the top 1\% cases of CAPES06 during March--August of 1980--2014 for each grid point, and the associated (b) median CAPE, (c) median S06, and (d) the residual, corresponding to each term in Eq. (7). Center of mass of the top 1\% cases of CAPES06 is represented by the product of the associated median CAPE and median S06. (e--h) as in (a--d) but for percentage changes in the center of the top 1\% cases of EHI03 and the associated terms, corresponding to Eq. (8).}
\label{fig_jointmap_notopo}
\end{figure}

Finally, we explicitly attribute the fractional changes in each combined proxy (CAPES06 or EHI03) to changes in their constituent parameters (CAPE, S06, or SRH03) using Eqs. (7) and (8). Specifically, we quantitatively examine responses in the center of mass of extreme CAPE-S06 or CAPE-SRH03 distribution and the contributions from changes in the associated median CAPE, S06, or SRH03. The center of mass is defined by the product of median CAPE and S06 or SRH03 associated with the top 1\% cases of CAPES06 or EHI03. In noTOPO, the relative contributions to decreases in extreme CAPES06 due to changes in CAPE and S06 varies across regions though both parameters are important (Figure \ref{fig_jointmap_notopo}a--d). Over the northern Great Plains and much of the eastern third of the U.S., the decrease ($\sim-40\%$) is driven more by a decrease in CAPE ($-20\%$--$-40\%$) than S06 ($\sim-20\%$). Over the southern Great Plains, the decrease ($-20\%$--$-40\%$) is driven primarily by a decrease in S06 ($-20\%$--$-40\%$). Decreases in extreme EHI03 are broadly similar to decreases in extreme CAPES06, though this response is driven more strongly by decreases in SRH03 rather than CAPE over much of the eastern half of the U.S. (Figure \ref{fig_jointmap_notopo}e--h). Hence, while the combined proxies appear to decrease broadly over eastern North America, the underlying reasons for their decrease actually vary regionally.

\begin{figure}[t]
\centerline{\includegraphics[width=19pc]{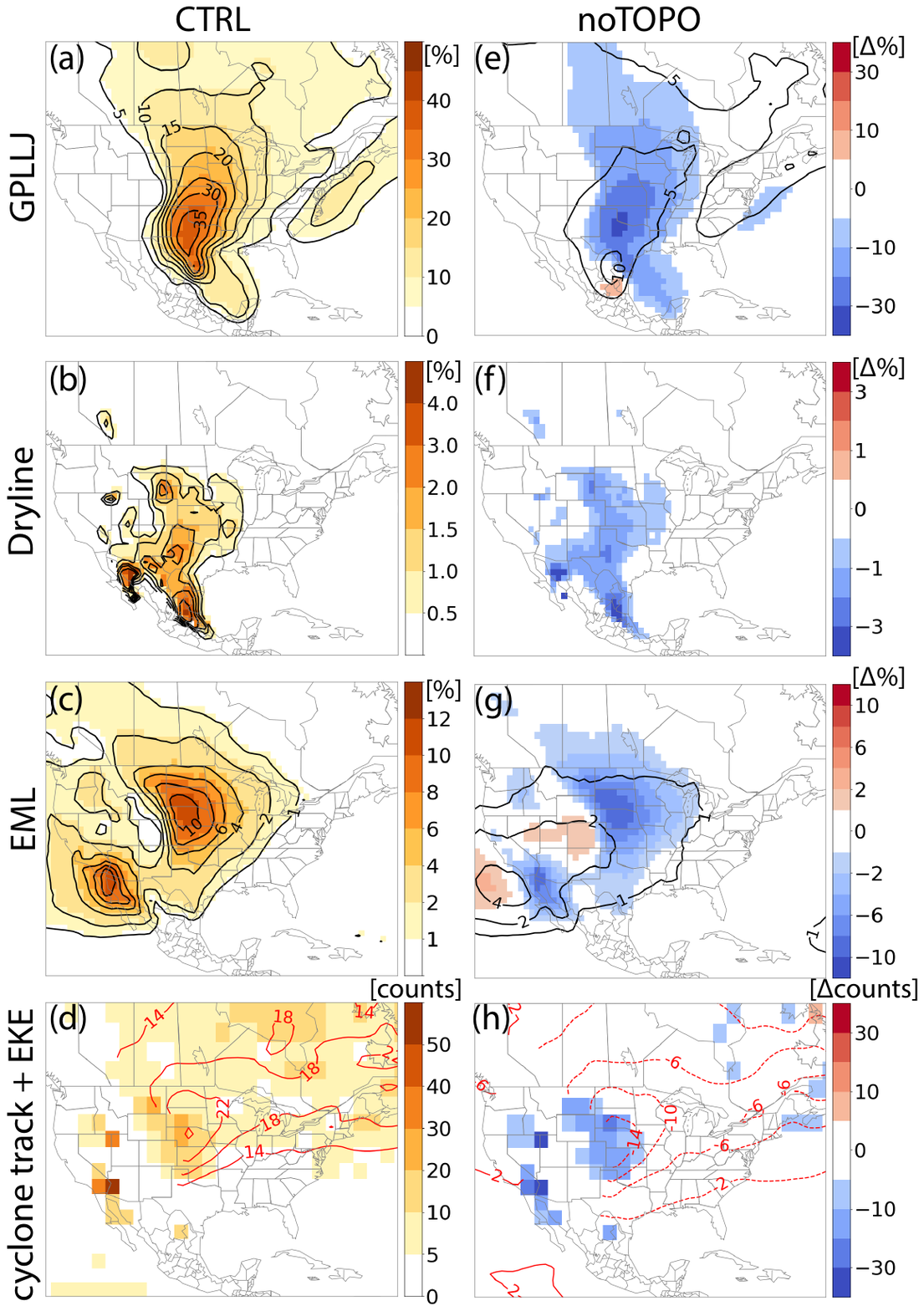}}
\caption{(a--c) CTRL percentage frequency (contour lines + filled contours) of southerly GPLLJs, drylines, and EMLs during March--August of 1980--2014; (d) CTRL mean frequency of cyclone track (counts per 2.5$^\circ$$\times$$2.5^\circ$ grid box; filled contours) and mean 2--6 day Butterworth bandpass filtered EKE at 850 hPa (m$^{2}$ s$^{-2}$, contour lines) during March--August of 1980--2014. (e--g) as in (a--c) but for noTOPO percentage frequency (contour lines) and the response (noTOPO percentages minus CTRL percentages; filled contours). (h) as in (d) but for noTOPO responses only.}
\label{fig_gpllj_notopo}
\end{figure}

\begin{figure*}[t]
\centerline{\includegraphics[width=32pc]{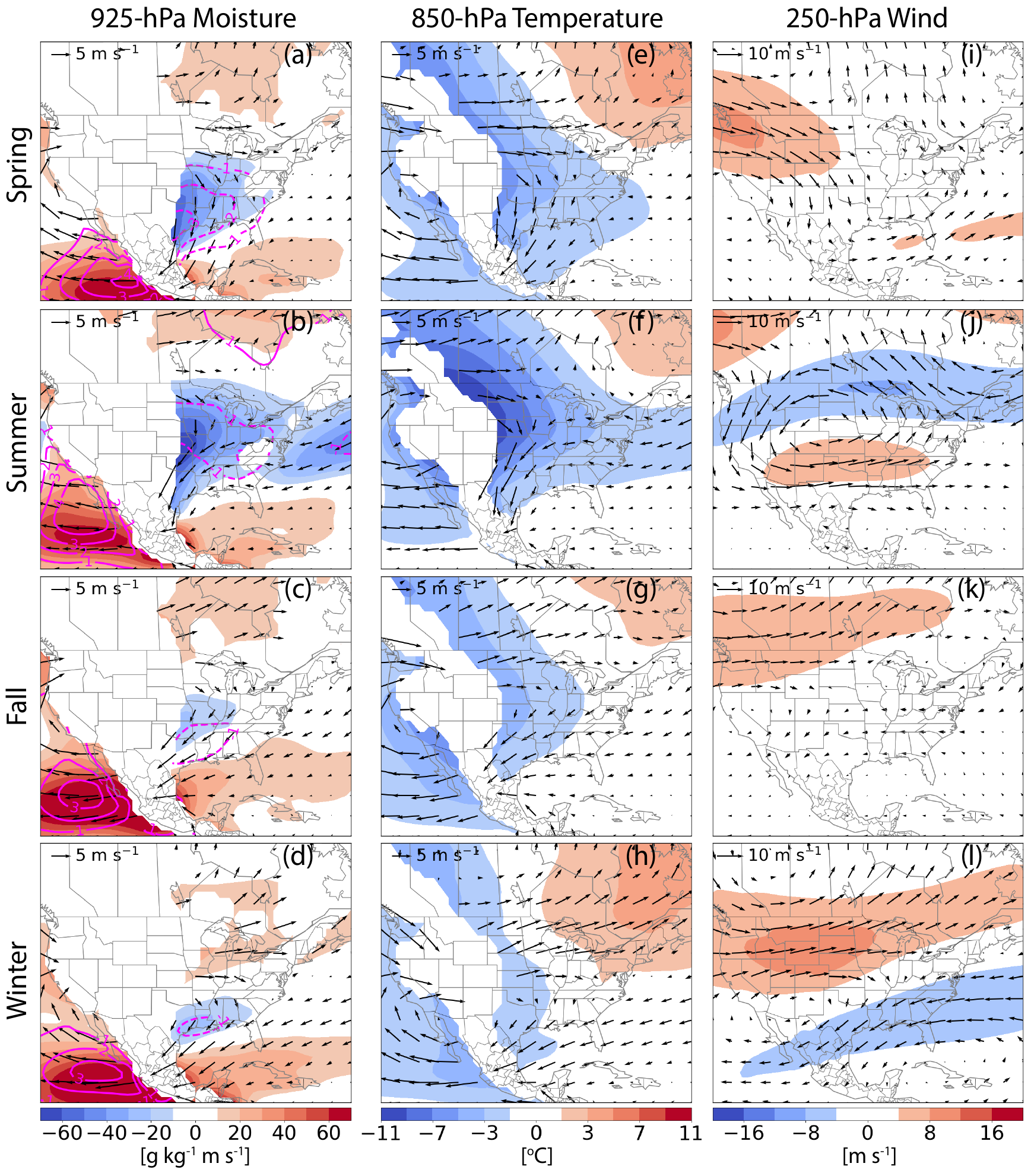}}
\caption{Responses of mean state in noTOPO (noTOPO minus CTRL) for each season (top--bottom: spring--winter). (a--d) specific humidity (g kg$^{-1}$; contour lines), wind vector (arrows), and horizontal moisture transport (filled contours) at 925 hPa. (e--h) air temperature (filled contours) and wind vector (arrows) at 850 hPa. (i--l) wind speed (filled contours) and wind vector (arrows) at 250 hPa.}
\label{fig_mean_notopo}
\end{figure*}

\begin{figure*}[t]
\centerline{\includegraphics[width=36pc]{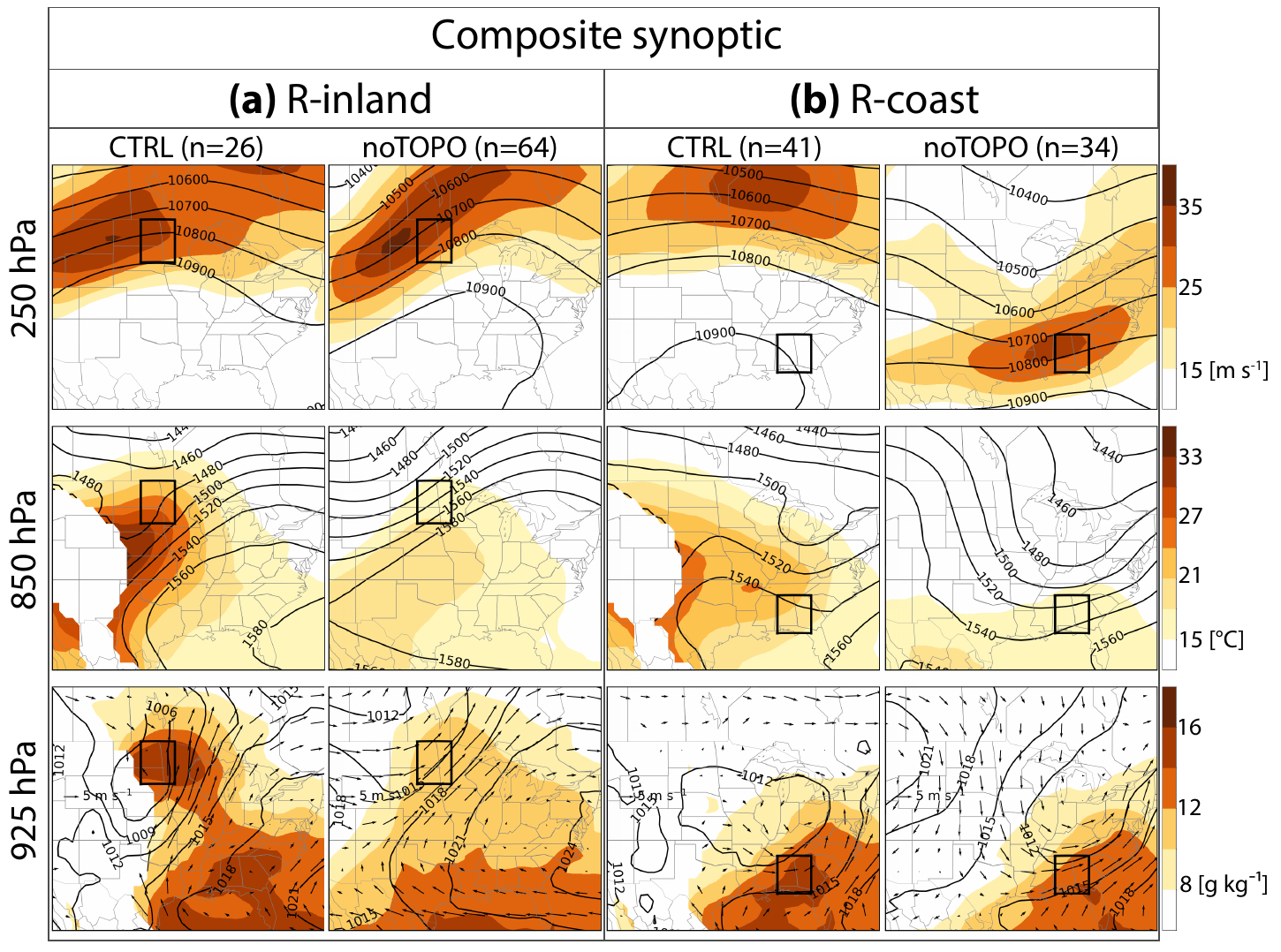}}
\caption{Composite synoptic patterns for cases associated with extreme SLS environments during March--August of 1980--2014 in (a) R-inland and (b) R-coast. Black squares indicate the location of sub-regions. Left column: CTRL. Right column: noTOPO. Top row: 250-hPa wind speed (filled contours) and geopotential height (m; contour lines). Middle row: 850-hPa temperature (filled contours) and geopotential height (m; contour lines). Bottom row: 925-hPa specific humidity (filled contours), wind vector (arrows), and sea level pressure (hPa; contour lines).}
\label{fig_composite_notopo}
\end{figure*}

\subsection{SLS-Relevant Synoptic-Scale Features} 

The substantial reduction of SLS environments described above is closely aligned with the responses of key synoptic-scale features commonly associated with SLS environments over North America. The southerly GPLLJs, drylines, EMLs, and extratropical cyclone activity are all strongly reduced during March--August (Figure \ref{fig_gpllj_notopo}), likely contributing to the less production of CAPE, S06 or/and SRH03 in noTOPO. 

Specifically, removing elevated terrain substantially reduces GPLLJs (Figure \ref{fig_gpllj_notopo}a,e), especially over the southern Great Plains. This response is qualitatively similar to past work \citep{pan2004,ting2006}. GPLLJs are not entirely eliminated by removing elevated terrain, though, as a local frequency maximum is still retained over northeastern Mexico (Figure \ref{fig_gpllj_notopo}e). Hence, upstream elevated terrain indeed appears essential to the production of GPLLJs, though land-ocean thermal contrast may also play a small role \citep{parish2000}. The reduced occurrence of GPLLJs indicates weakened mean low-level meridional winds, which thus contributes to the slight weakening of S06 in summer and the strong weakening of SRH03 in spring and summer. Moreover, it may correspond to reduced northward moisture transport into the continental interior, which may partially explain the broad reduction of CAPE in noTOPO; this topic is examined further below. 

Removing elevated terrain effectively eliminates drylines (Figure \ref{fig_gpllj_notopo}b,f). This implies a weakening of horizontal near-surface moisture gradients and moisture convergence over the Great Plains. This outcome reflects both the enhanced surface moisture to the west by the reduced elevation and the decreased surface moisture over the Great Plains by the weakened northward transport of moisture from the Gulf of Mexico. 

Removing elevated terrain strongly reduces EMLs over the Great Plains, though does not eliminate them (Figure \ref{fig_gpllj_notopo}c,g). Though perhaps unsurprising, these results confirm that the elevated terrain to the west are indeed essential to the generation of EMLs east of these mountains \citep{Carlson_etal_1983}. Since the presence of an EML and the associated capping inversion permit the build-up of CAPE in an atmospheric column by inhibiting convective initiation \citep{Carlson_etal_1983,Farrell_Carlson_1989,Banacos_Ekster_2010, Ribeiro_Bosart_2018}, this reduction of EMLs is fully consistent with the reduction of CAPE and CIN found above.

Finally, removing elevated terrain also reduces extratropical cyclone activity substantially, especially on the lee of the Rocky Mountains over the central U.S. where the local maximum of cyclone track and 850-hPa EKE is almost eliminated (Figure \ref{fig_gpllj_notopo}d,h). These results indicate the key role of the Rocky Mountains in generating stationary waves and thus localizing extratropical cyclone activity over east-central North America, in line with past studies \citep{broccoli1992,inatsu2002,brayshaw2009}. Given the east-northeastward storm track (Figure \ref{fig_gpllj_notopo}d) \citep{reitan_1974, zishka_1980}, a reduction of cyclogenesis on the lee of the Rocky Mountains further reduces cyclone activity over the Great Lakes and off the northeastern U.S. coast (Figure \ref{fig_gpllj_notopo}h). Fewer extratropical cyclones also likely contribute to the inland reduction in CAPE due to the weakened low-level moisture and heat convergence associated with the warm sectors of extratropical cyclones \citep{Hamill_etal_2005, Tochimoto_etal_2015}. Moreover, their absence will also reduce S06 and SRH03 typically generated by the cyclonic circulation and associated with these baroclinic systems \citep{Doswell_Bosart_2001}. Note though that removing elevated terrain has a weaker influence on tracked cyclone activity over higher latitudes in Canada where the jet stream remains relatively zonal.

\subsection{Mean State and Characteristic Synoptic Flow Pattern}

To further understand these responses of SLS environments, we analyze responses in the mean synoptic patterns at lower and upper levels in noTOPO (Figure \ref{fig_mean_notopo}). Removing elevated terrain substantially dries and cools the troposphere over much of the eastern half of the U.S. (Figure \ref{fig_mean_notopo}a--h), especially at low--mid levels in spring and summer (Figure \ref{fig_mean_notopo}a,b,e,f). This response is consistent with the strong reduction of CAPE and the associated synoptic-scale features. Specifically, the drying response throughout the regions, especially over the Great Plains, is associated with the strong weakening in southerly winds at 925 hPa (and 850 hPa) and hence the reduced GPLLJs, which substantially reduces northward moisture transport from the Gulf of Mexico (Figure \ref{fig_mean_notopo}a--d). Meanwhile, the Great Plains exhibit a strong cooling at 850 hPa that is consistent with the strong reduction of EMLs and hence of CAPE and CIN (Figure \ref{fig_mean_notopo}e--h). This cooling response weakens moving upward (not shown), and thus reduces mid-level lapse rates, indicative of the CAPE reduction as well. The horizontal flow becomes more zonal at all levels; the substantially weakened low-level meridional winds in spring and summer and the slightly weakened upper-level jet streams in summer (Figure \ref{fig_mean_notopo}j) are consistent with the reductions of S06 and SRH03 in warm seasons. Note that the mean upper-level jet streams are enhanced in winter and spring over much of the northern half of the U.S. (Figure \ref{fig_mean_notopo}l), which likely contributes to the enhanced S06 during these months over this region. 

These responses in the mean-state flow are also found in the composite synoptics of extreme SLS environments (Figure \ref{fig_composite_notopo}). Overall, the composite flow pattern is similar without topography in the inland region (R-inland) (Figure \ref{fig_composite_notopo}a), whereas it differs markedly for the coastal region (R-coast) (Figure \ref{fig_composite_notopo}b). For R-inland, the region is located downstream of a trough at all levels that advects warm and moist air northward into the continental interior  (Figure \ref{fig_composite_notopo}a) with an upper-level jet stream to the west in each. The primary difference is the weakened warm advection from the upstream elevated terrain at 850 hPa and the weakened southerly winds and moisture advection from the Gulf of Mexico at 925 hPa, associated with a weaker surface cyclone in noTOPO; this indicates that the cooling and drying response in the mean state atmosphere discussed above persists in composite synoptics of the extreme cases. For R-coast, the low-level pattern of warm and moist advection remains in noTOPO (Figure \ref{fig_composite_notopo}b), but at mid- and upper-levels a deep trough exists upstream, which differs markedly from the ridge pattern found in CTRL particularly at 250 hPa. Hence, without elevated terrain, the characteristic synoptic flow pattern is qualitatively similar between the continental interior and near the Gulf Coast, whereas they are qualitatively different in CTRL. This distinct behavior in synoptic set-up may be relevant to the reduced predictability of SLS activity in the Southeast \citep{Miller_mote_2017}.

\section{Results: Responses to filling in the Gulf of Mexico (noGOM)}\label{sec:results-nogom}

Following the structure of Section \ref{sec:results-notopo}, we next analyze responses to the removal of the Gulf of Mexico by comparing noGOM with CTRL. 

\subsection{SLS Environments} 

\begin{figure}[t]
\centerline{\includegraphics[width=19pc]{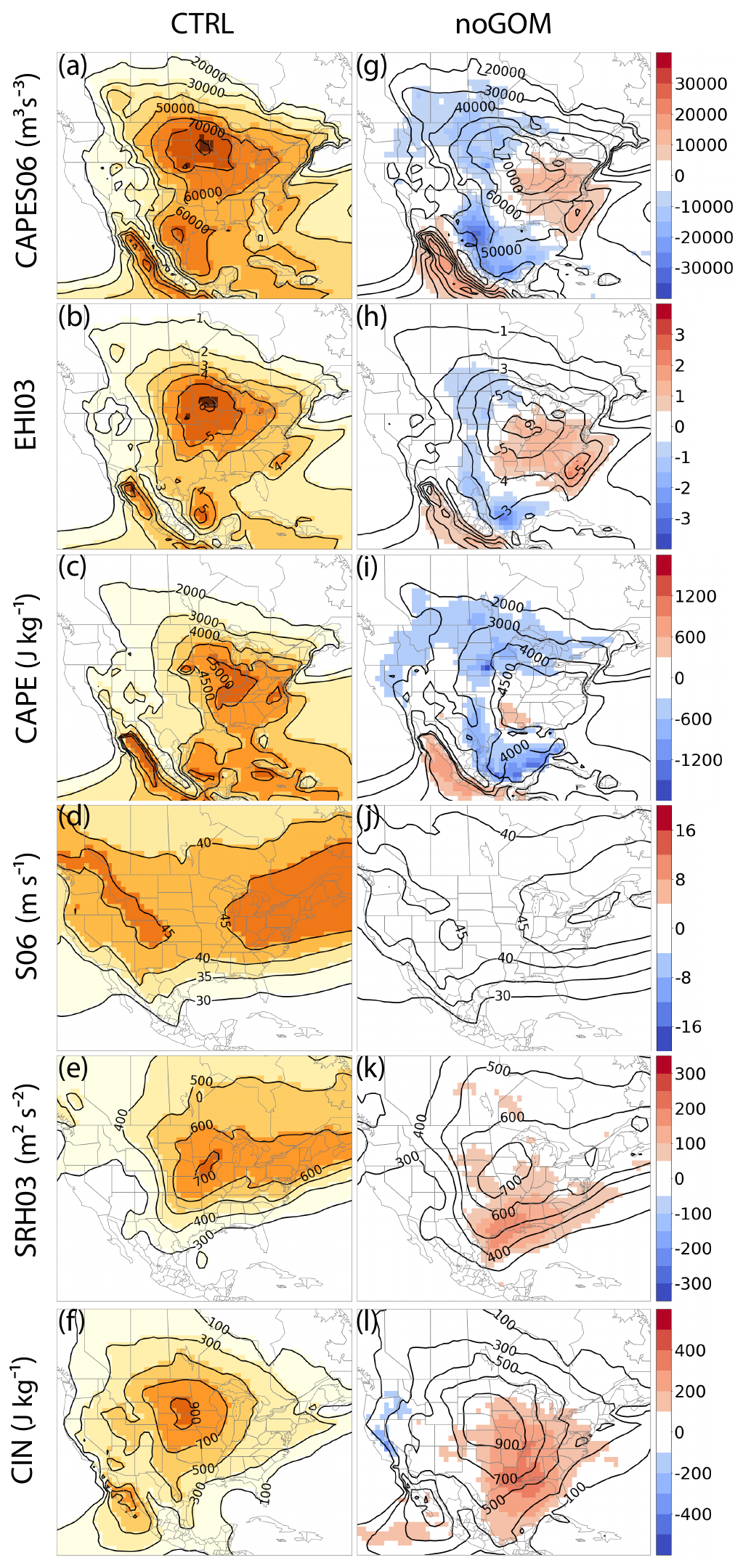}}
\caption{CTRL vs. noGOM for the 99th percentiles of CAPES06, EHI03, CAPE, S06, SRH03, and CIN during March--August of 1980--2014. Plot aesthetics as in Fig. \ref{fig_99ile_annual_notopo}.}
\label{fig_99ile_annual_nogom}
\end{figure}

In contrast to noTOPO, replacing the Gulf of Mexico with land does not strongly change the overall amplitude of extreme CAPES06 and EHI03 over North America relative to CTRL, though it does induce changes in the spatial pattern (Figure \ref{fig_99ile_annual_nogom}a,b,g,h). The primary local maximum shifts southeastward to the Midwest centered over Illinois. This shift emerges due to the reduction of SLS environments over the northern Great Plains (CAPES06: $-10,000$ m$^{3}$ s$^{-3}$; EHI03: $-1$) and the enhancement over the eastern third of the U.S. (CAPES06: $+10,000$ m$^{3}$ s$^{-3}$; EHI03: $+$1). Meanwhile, the secondary local maximum over southern Texas is eliminated, as CAPES06 and EHI03 are substantially reduced over the region in noGOM ($-20,000$ m$^{3}$ s$^{-3}$ and $-1.5$). For constituent parameters, CAPE decreases ($\sim-600$ J kg$^{-1}$) over the Great Plains (Figure \ref{fig_99ile_annual_nogom}c,i), whereas S06 and SRH03 change only minimally (Figure \ref{fig_99ile_annual_nogom}d,e,j,k). SRH03 is slightly enhanced ($\sim+100$ m$^{2}$ s$^{-2}$) over the south-central U.S. Finally, CIN is enhanced over the Gulf of Mexico itself and much of the eastern half of the U.S. extending inland from the Gulf of Mexico (Figure \ref{fig_99ile_annual_nogom}f,l). Note that CAPE changes relatively little in this region, a behavior consistent with drier and warmer boundary-layer air at roughly fixed moist static energy in this region (Figure S2), which will increase CIN while keeping CAPE relatively constant \citep{Agard_Emanuel_2017, chen2020, chavas_dawson_2020, taszarek2020_03}.

\begin{figure*}[t]
\centerline{\includegraphics[width=36pc]{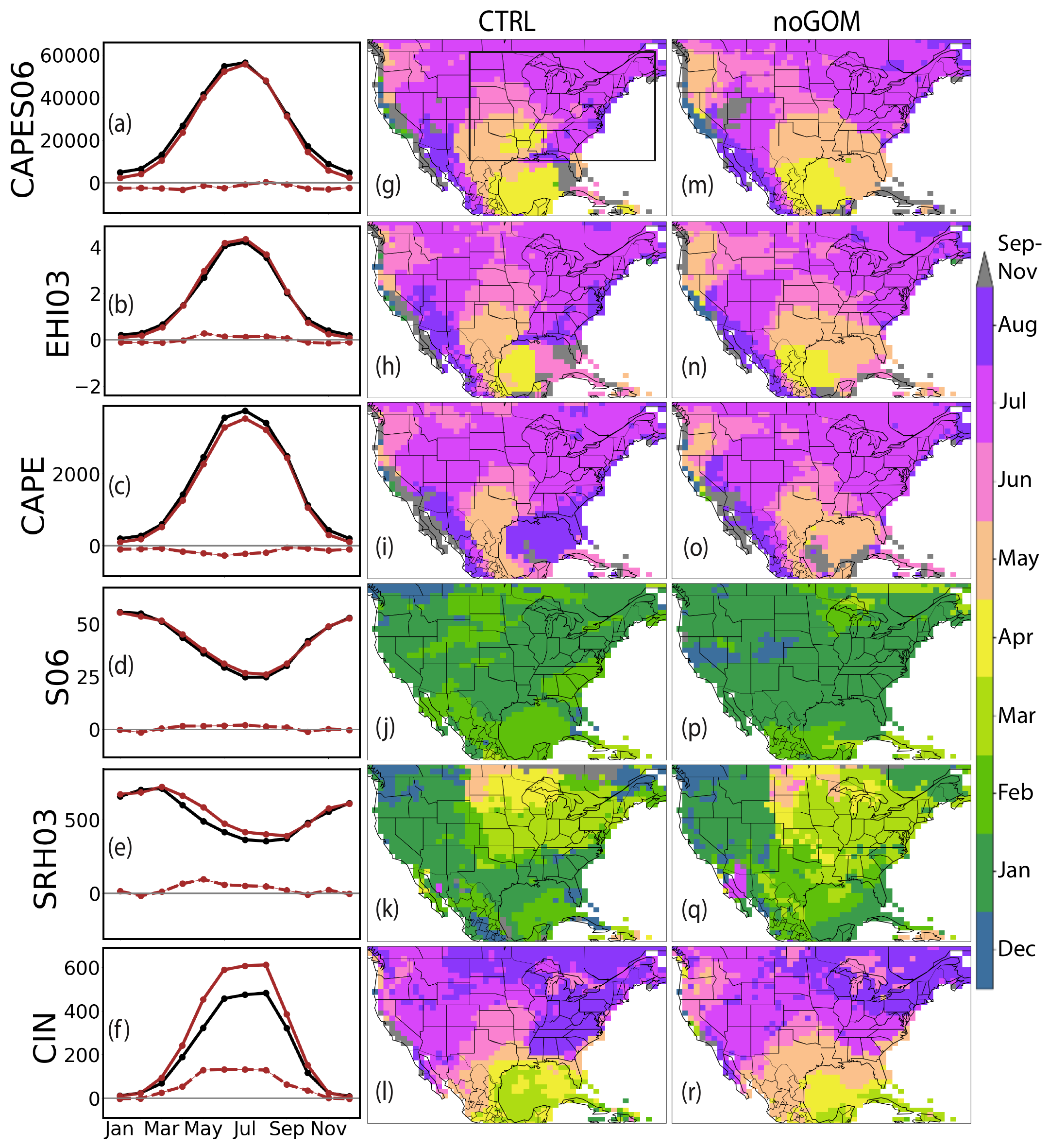}}
\caption{CTRL vs. noGOM for seasonal cycles of extreme CAPES06, EHI03, CAPE, S06, SRH03, and CIN. Plot aesthetics as in Fig. \ref{fig_99ile_peakmm_notopo}.}
\label{fig_99ile_peakmm_nogom}
\end{figure*}

These responses are consistent throughout the year and thus the amplitudes of their seasonal cycles remain relatively constant (Figure \ref{fig_99ile_peakmm_nogom}a--f). Nonetheless, replacing the Gulf of Mexico with land does indeed slightly reduce CAPE and increase S06 and SRH03 in spring and summer (Figure \ref{fig_99ile_peakmm_nogom}c--e). These responses offset each other in the seasonal cycle of CAPES06 and EHI03 resulting in minimal changes in each (Figure \ref{fig_99ile_peakmm_nogom}a,b). The only exception is CIN, whose seasonal cycle amplitude is enhanced due to increases in the warm seasons (Figure \ref{fig_99ile_peakmm_nogom}f).

In terms of their seasonal phase, replacing the Gulf of Mexico with land causes seasonal phase shifts primarily over the southeastern U.S., as well as over the Gulf of Mexico region itself (Figure \ref{fig_99ile_peakmm_nogom}g--r). Specifically, CAPES06 and EHI03 peak 2--3 months earlier over the southeastern U.S., shifting from July or August in CTRL to May in noGOM (Figure \ref{fig_99ile_peakmm_nogom}g,h,m,n). The peak month of CAPE remains broadly similar to CTRL (Figure \ref{fig_99ile_peakmm_nogom}i,o), though it peaks earlier over the Gulf of Mexico (May in noGOM; August in CTRL). Meanwhile, the peak month in S06 shifts slightly earlier over the southeastern U.S. and the Gulf of Mexico from February to January, whereas SRH03 over the Gulf coast shifts slightly later to February or March from January (Figure \ref{fig_99ile_peakmm_nogom}j,k,p,q). Finally, CIN peaks 1 month earlier over the southeastern U.S, shifting from August in CTRL to July in noGOM (Figure \ref{fig_99ile_peakmm_nogom}l,r). Overall, replacing the Gulf of Mexico with land extends the inland progression of the seasonal cycle over the Great Plains farther east into the southeastern U.S., thereby reducing (though not eliminating) zonal variability in the seasonal cycle, similar to noTOPO.

\begin{figure}[t]
\centerline{\includegraphics[width=19pc]{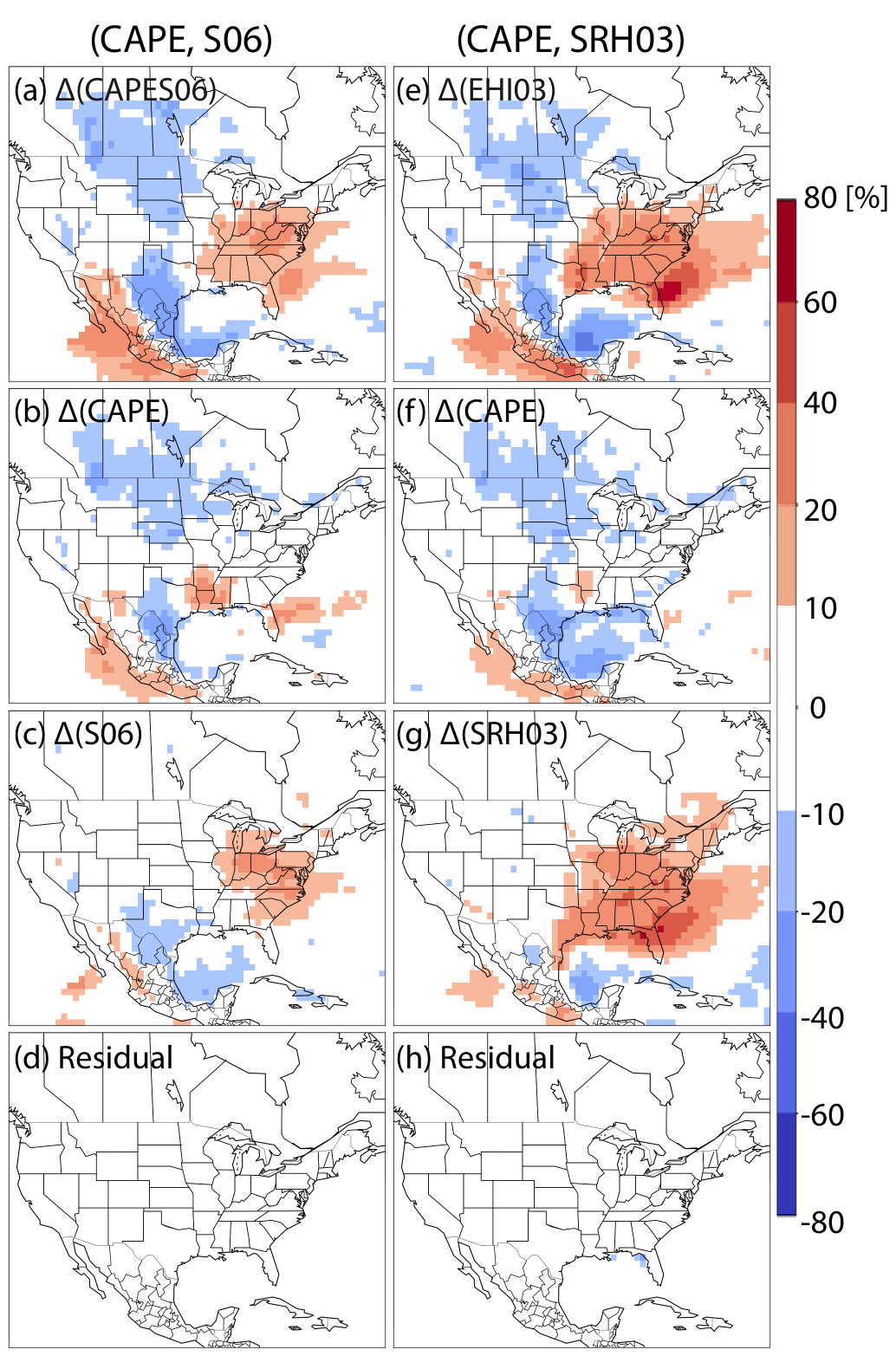}}
\caption{Attributions of responses in noGOM extreme CAPES06 and EHI03. Plot aesthetics as in Fig. \ref{fig_jointmap_notopo}.}
\label{fig_jointmap_nogom}
\end{figure}

Similarly, we explicitly attribute the changes in each combined proxy (CAPES06 or EHI03) to changes in their constituent parameters (CAPE, S06, or SRH03) using Eqs. (7) and (8) (Figure \ref{fig_jointmap_nogom}). In noGOM, CAPES06 and EHI03 change more subtly and non-uniformly in space. Specifically, the decrease of CAPES06 and EHI03 over the northern Great Plains is driven predominantly by a decrease in CAPE, while their increase over much of the eastern U.S. is primarily driven by an increase in S06 and SRH03. In combination, the result is an eastward shift of the primary local maximum of CAPES06 and EHI03. Meanwhile, the removal of the secondary local maximum of CAPES06 and EHI03 over southern Texas is driven by decreases in both CAPE and S06 (and SRH03), though CAPE contributes a larger portion. 

\subsection{SLS-Relevant Synoptic-Scale Features} 

\begin{figure}[t]
\centerline{\includegraphics[width=19pc]{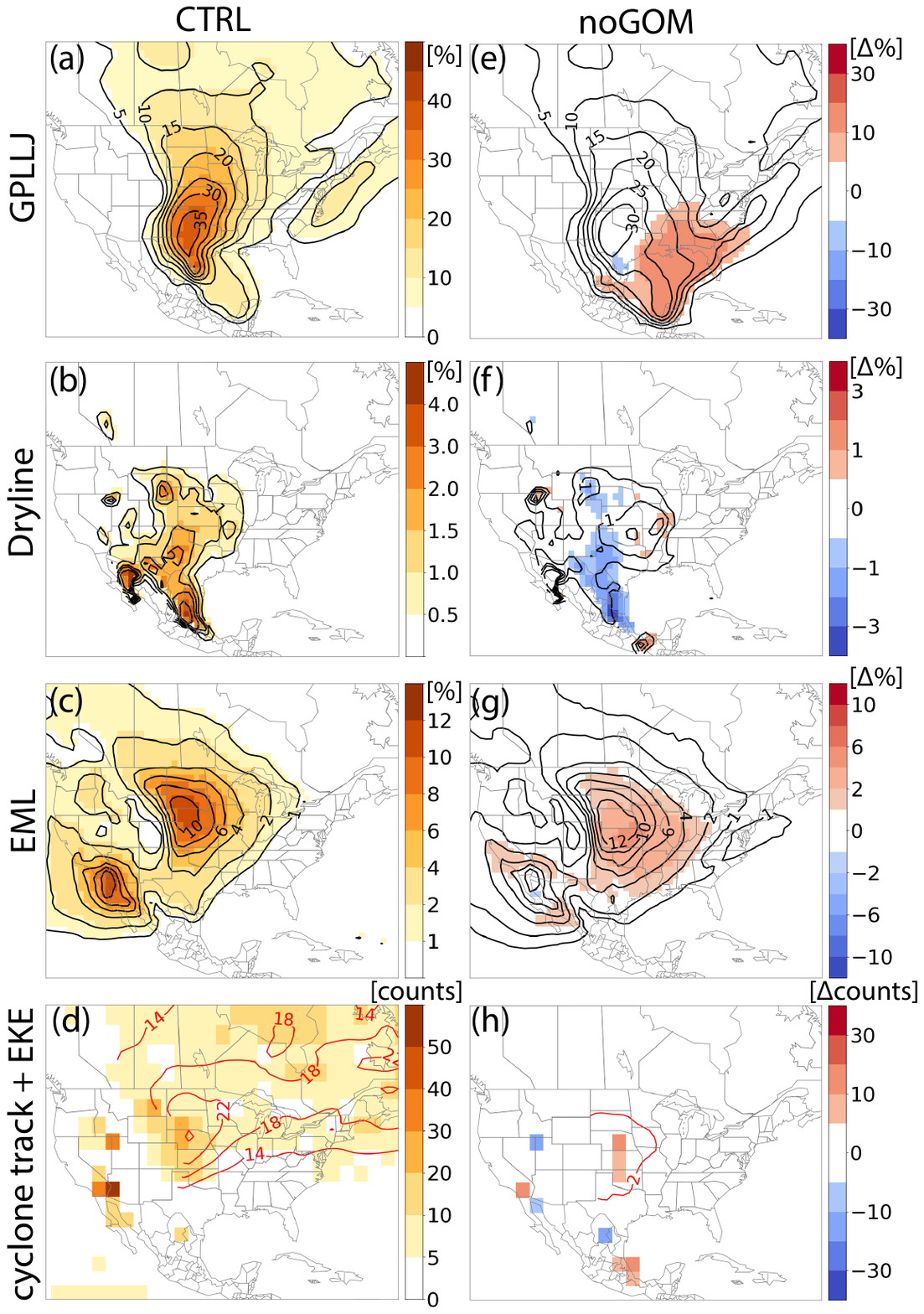}}
\caption{CTRL vs. noGOM for the March--August frequency of southerly GPLLJs, drylines, EMLs, and extratropical cyclone activity. Plot aesthetics as in Fig. \ref{fig_gpllj_notopo}.}
\label{fig_gpllj_nogom}
\end{figure}

These SLS environmental responses in noGOM are again consistent with the muted responses of the associated synoptic-scale drivers over North America, as the southerly GPLLJs, drylines, EMLs, and extratropical cyclone activity all alter minimally in spring and summer (Figure \ref{fig_gpllj_nogom}).

Specifically, replacing the Gulf of Mexico with land has little influence on the frequency of GPLLJs, which is maximized over the southern Great Plains (Figure \ref{fig_gpllj_nogom}a,e). Yet the presence of the new land still acts to reduce the local moisture supply at low levels over the Gulf of Mexico, and thus potentially reduces northward moisture transport into the Great Plains (detailed below). This may contribute to the reduction of CAPE in noGOM over the Great Plains found above. In addition, noGOM does produce an increase in GPLLJs over the Gulf of Mexico and the deep south (Figure \ref{fig_gpllj_nogom}e), which are likely a combination of topographicaly-forced (mainly the Mexican Plateau) GPLLJs and coastal GPLLJs due to the enhanced land-ocean thermal contrast between the new land and tropical ocean \citep{parish2000}. The increased frequency of GPLLJs indicates enhanced mean low-level winds, which may explain the enhanced S06 and SRH03 associated with CAPES06 and EHI03 over much of the eastern third of the U.S. 

Replacing the Gulf of Mexico with land nearly eliminates drylines over the southern Great Plains and northeastern Mexico consistent with the reduced CAPE there, while it has little influence on drylines over the central U.S. (Figure \ref{fig_gpllj_nogom}b,f). Note that these reduced drylines primarily form in spring, while the retained drylines occur mainly in summer (not shown). These results indicate that the Gulf of Mexico serves as an essential source of moisture for the southern Great Plains in spring that favors dryline formations. In contrast, in summer other sources of moisture (e.g., soil and vegetation) likely contribute strongly to dryline formations, in line with findings of \citet{Molina_etal_2019}. 

Replacing the Gulf of Mexico with land produces slightly more EMLs over the central U.S. (Figure \ref{fig_gpllj_nogom}c,g). Considering that the horizontal advection of elevated air masses with steep lapse rates dominates formation and maintenance of EMLs over North America \citep{Banacos_Ekster_2010, Ribeiro_Bosart_2018}, the increase in EMLs is in part consistent with the enhanced CIN, though it does not necessarily increase CAPE.  

Finally, replacing the Gulf of Mexico with land has little influence on extratropical cyclone activity, as the cyclone track frequency and 850-hPa EKE in noGOM are broadly similar to CTRL, though with a slight increase over a small region in the central Great Plains (Figure \ref{fig_gpllj_nogom}d,h).

\begin{figure*}[t]
\centerline{\includegraphics[width=32pc]{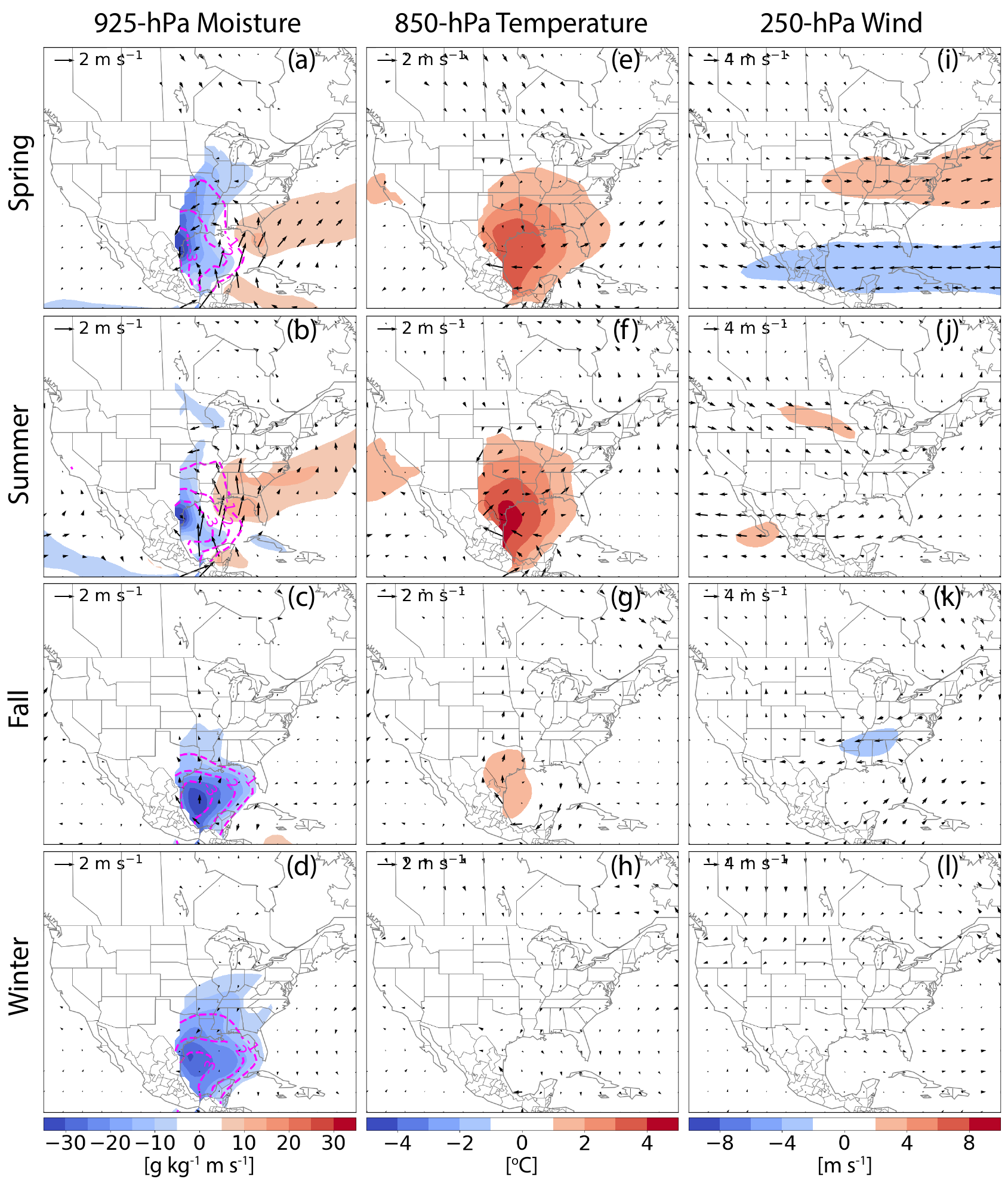}}
\caption{The noGOM responses of seasonal mean state at 925, 850, and 250 hPa. Plot aesthetics as in Fig. \ref{fig_mean_notopo}.}
\label{fig_mean_nogom}
\end{figure*}

\begin{figure*}[t]
\centerline{\includegraphics[width=36pc]{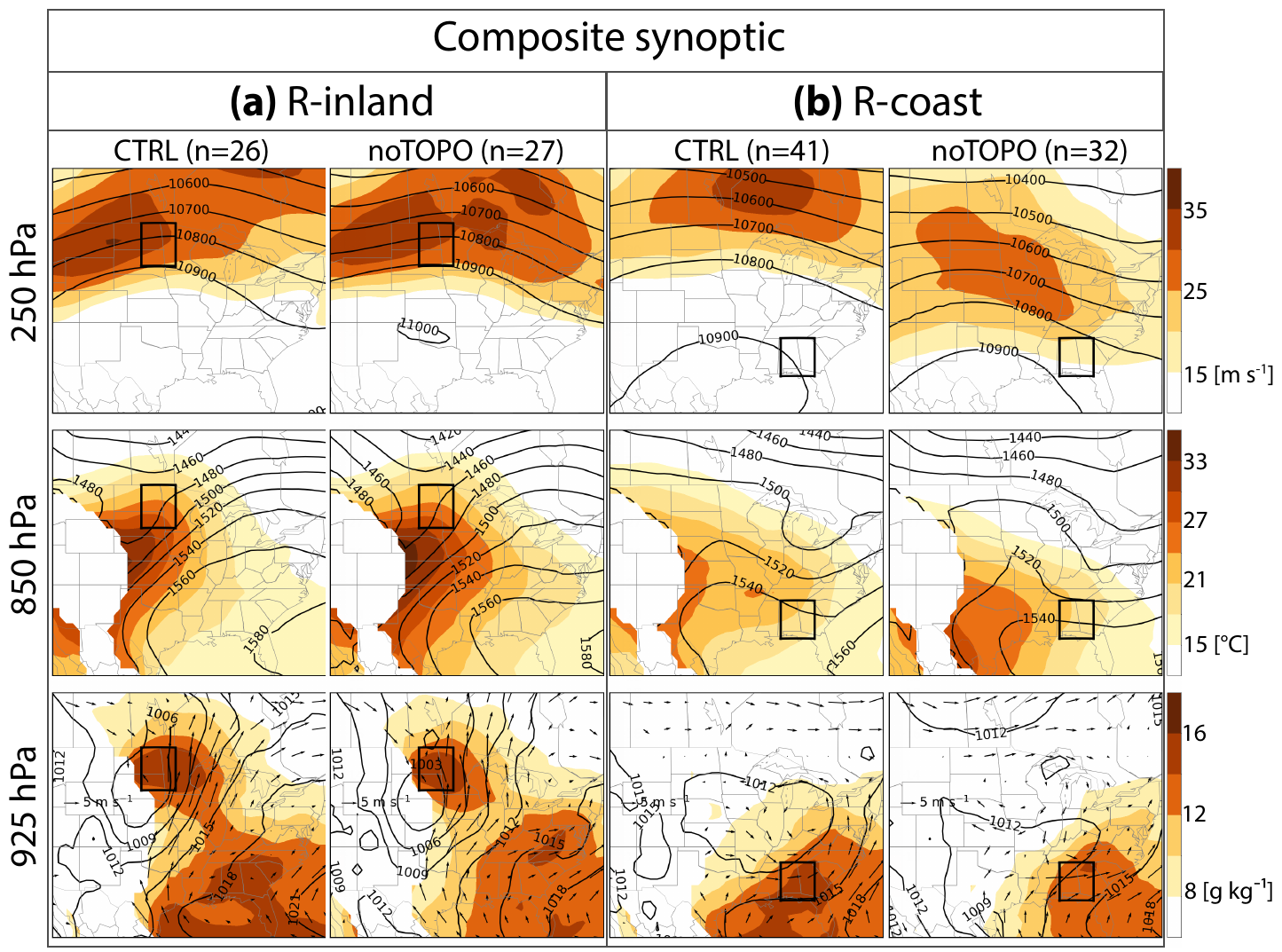}}
\caption{CTRL vs. noGOM for the composite synoptic patterns of extreme SLS environments. Plot aesthetics as in Fig. \ref{fig_composite_notopo}.}
\label{fig_composite_nogom}
\end{figure*}

\subsection{Mean State and Characteristic Synoptic Flow Pattern}

Finally, to further understand the responses of SLS environments, we analyze responses in the mean synoptic patterns at lower and upper levels in noGOM (Figure \ref{fig_mean_nogom}). Replacing the Gulf of Mexico with land causes the mean-state response to dry and warm at low levels. This response is strongest over the Gulf of Mexico region itself and southern Texas (Figure \ref{fig_mean_nogom}a--h), due to the reduced surface moisture supply from the new land and its lower heat capacity as compared to ocean water. These responses over the Gulf of Mexico strongly affect the the southern Great Plains via inland advection of this drier and warmer air by the large-scale flow. While the enhanced 850-hPa temperature in spring and summer over the southern Great Plains ($\sim+3$ K) may actually contribute to the modest increase in the frequency of EMLs over the Great Plains noted above, which could translate to higher CAPE in isolation, this effect is offset by the low-level drying response ($\sim-2$ g kg$^{-1}$), resulting in a derease in CAPE; this is further supported by the reduction of low-level moist static energy over the Great Plains (Figure S2). 

Ultimately, this lower-tropospheric drying and warming effect moves farther inland toward the continental interior though weakens in magnitude (Figure \ref{fig_mean_nogom}a--h). Meanwhile, there is a slight enhancement of moisture transport at 925 hPa in spring and summer over the east Gulf coast (Figure \ref{fig_mean_nogom}a,b), possibly due to the strengthened southerly winds (and hence the increased GPLLJs) enhancing moisture transport from the remote tropical ocean. Analysis of changes in the associated moisture budget could quantify this effect \citep{Molina_etal_2019,Molina_etal_2020} in future work. Though the drying effect dominates the reduced CAPE over the Great Plains, the warming response itself does indeed increase low-level dry static energy over much of the eastern U.S., and thus contributes to the enhanced CIN found above. In addition, the inland flow remains strong in spring and summer when persistent low-level southerly winds are present (Figure \ref{fig_mean_nogom}a--d). These southerly winds are enhanced over the deep south, especially in summer, consistent with the increased GPLLJs noted above. This flow response is also consistent with the enhanced S06 and SRH03 associated with CAPES06 and EHI03 over much of the eastern third of the U.S. The upper tropospheric flow in noGOM is broadly similar to CTRL (Figure \ref{fig_mean_nogom}i--l), indicating that the mean-state response to filling the Gulf of Mexico is confined primarily to the lower troposphere.

The composite synoptic patterns of extreme SLS environments have relatively small differences between noGOM and CTRL in both regions R-inland and R-coast (Figures \ref{fig_composite_nogom}a,b). The 850-hPa temperature increases slightly in noGOM (Figures \ref{fig_composite_nogom}a,b), indicative of the low-level warming effects from the Gulf of Mexico as was found in the mean-state response. The 925-hPa moisture and flow fields do not change strongly in noGOM, indicating that local moisture sources (e.g., soil and vegetation) or far-field sources from the tropical oceans are more important for the formation of extreme SLS environments over eastern North America, in line with \cite{Molina_etal_2019}. In contrast to responses in noTOPO, replacing the Gulf of Mexico does not impact differences in the composite synoptic between R-inland and R-coast. Hence, it confirms that the regional variability in characteristic synoptic flow patterns between the Great Plains and the Southeast primarily depends on the upstream elevated terrain rather than the Gulf of Mexico.  

\section{Conclusions}\label{sec:conclusions}

The eastern half of North America is one of the principal hot spots for SLS activity globally. The prevailing conceptual model for this behavior, proposed by \cite{Carlson_etal_1983}, posits that elevated terrain to the west and the Gulf of Mexico to the south together are critical to produce environments conducive to SLS activity. We test this conceptual model in the CAM6 global climate model by conducting two experiments relative to an AMIP-style control run (CTRL): 1) North American topography removed (noTOPO), and 2) the Gulf of Mexico converted to land (noGOM). We focused our analysis on responses of SLS environments during warm seasons, defined by extreme values (99th percentiles) of two common SLS environmental proxies (CAPES06 and EHI03) during March--August, and quantitatively attributed their changes to changes in their constituent parameters (CAPE, S06, and SRH03). To better understand these responses, we next analyzed responses of a set of key synoptic-scale features commonly associated with the generation of SLS environments: southerly GPLLJs, drylines, EMLs, and extratropical cyclone activity. Finally, we analyzed changes in the mean state and characteristic synoptic flow patterns to understand how these responses are related to changes in the large-scale circulation. 

We summarize our primary results as follows:

\begin{enumerate}

\item The existence of SLS environments over North America does indeed depend strongly on upstream elevated terrain (noTOPO vs. CTRL).

    \begin{enumerate}
    
    \item Removing elevated terrain substantially reduces extreme CAPES06 and EHI03 over North America, particularly over the northern Great Plains, leaving a residual distribution that is maximized near the Gulf coast and decays moving inland towards the continental interior. Their reduction is driven by the reduction of all three constituent parameters (CAPE, S06 and SRH03), though their contributions vary spatially. The response of these SLS proxies occurs largely during peak SLS seasons in spring and summer, and thus reduces the amplitude of their seasonal cycles. Moreover, removing elevated terrain suppresses the inland progression of the seasonal cycle and leads to relatively uniform seasonal phase that peaks in June over the southern U.S. 
    
    \item This response is accompanied by a strong reduction in the occurrence of all key SLS-relevant synoptic-scale features. The reduced EMLs indicate the weakened downstream advection of warm well-mixed layers with steep lapse rates from the elevated terrain, consistent with the cooling of the 850 hPa mean state, especially over the Great Plains. The reduced GPLLJs and extratropical cyclones are consistent with reduced inland low-level moisture transport from the Gulf of Mexico which dries the lower troposphere, as well as a more zonal tropospheric mean flow. Thus, taken together, the cooler and drier mean-state atmosphere with weakened low-level inland winds is less favorable to the generation of CAPE, S06 and SRH03, and hence the formation of SLS environments, despite the reduction in CIN. The characteristic synoptic flow pattern associated with SLS environments near the Gulf coast differs markedly from the classic upstream trough pattern found inland in CTRL, but it becomes similar when elevated terrain is removed, indicating that elevated terrain generates spatial variability in how SLS envrionments are produced by the large-scale flow.  
    \end{enumerate}

\item The existence of SLS environments over North America depends much less strongly on the Gulf of Mexico though its effects are not negligible (noGOM vs. CTRL). 

    \begin{enumerate}
    
    \item Replacing the Gulf of Mexico with land shifts the primary local maximum of extreme CAPES06 and EHI03 southeastward from the northern Great Plains into the southern Midwest, primarily driven by a reduction of CAPE over the northern Great Plains and an increase in S06 and SRH03 over the eastern third of the U.S. It further eliminates the secondary, smaller local maximum over the southern Great Plains, primarily driven by the reduction of CAPE. The amplitude of the seasonal cycle is not strongly influenced, though its spatial footprint expands eastward, with its peak over the Southeast U.S. shifting 1--2 months earlier to May similar to the southern Great Plains. 
    
    \item Consistent with modest responses in SLS environments, there are modest changes in the key SLS-relevant synoptic-scale features. Drylines are reduced in spring, consistent with the drying of the 925 hPa mean state, resulting in a decrease in CAPE over the Great Plains. The enhanced EMLs are consistent with the strong warming of the 850 hPa mean state, which may contribute to the increase in CIN though it does not necessarily translate to increasing CAPE. This lower tropospheric drying and warming response in mean state is the strongest near the west Gulf coast and decreases in magnitude moving inland. Meanwhile, GPLLJs are increased over the deep south that is consistent with the stronger low-level meridional winds, and thus partly contribute to the enhanced S06 and SRH03. Ultimately, the contrast response in CAPE and S06 or SRH03 that varies regionally induces the subtle changes in SLS environments. The characteristic synoptic flow patterns that generate extreme SLS environments do not strongly depend on the Gulf of Mexico.
    
    \end{enumerate}

\end{enumerate}

We also conducted an additional experiment with both North American topography removed and the Gulf of Mexico converted to land (Figure S3). Responses in this experiment are broadly similar to responses in noTOPO, confirming that the presence of elevated terrain plays a critical role in producing downstream SLS environments as found over present-day North America, whereas the Gulf of Mexico plays a secondary role. The removal of these geographic components still leaves a residual peak of SLS environments near the southeast coast that decays inland and hence appears strongly driven by land-ocean contrast. The presence of the Gulf of Mexico shifts the primary local maximum of SLS environments westward closer to the elevated terrain, and acts as the essential moisture source for producing the secondary local maximum of these environments over the southern Great Plains. Thus, changes over the Gulf of Mexico (e.g., SST) may alter the spatial distribution of SLS environments; further investigation may provide insight into understanding the eastward shift of these environments observed in recent decades \citep{Gensini_Brooks_2018, tang2019}. Meanwhile, as noted earlier, though we filled the Gulf of Mexico with grassland in noGOM for simplicity, different land types may affect the details of these responses. Moreover, here we removed any elevated terrain over all of North America in noTOPO, but specific topographic features, such as the Appalachian Mountains, may induce more localized responses; this could be a valuable avenue for future work. Note that extreme CAPES06 and EHI03 as defined here primarily represent high CAPE cases in spring and summer and hence are less representative of high shear low CAPE environments that are more common in the cool seasons in the southeast U.S. \citep{guyer_2010,sherburn_etal_2014,Sherburn_etal_2016,li2020}. Removing elevated terrain does lead to an increase in S06 during the winter, which may yield different responses to changes in elevated terrain on SLS environments in the cold seasons. Finally, in addition to experiments with real-Earth global climate models, idealized models with simplified settings could provide a more robust testing ground for understanding how surface properties control the formation of SLS environments on Earth.

\acknowledgments{} 

We thank Editor Dr. Xin-Zhong Liang and three anonymous reviewers for their feedback in improving this manuscript. We would like to acknowledge high-performance computing support from Cheyenne (doi:10.5065/D6RX99HX) provided by NCAR's Computational and Information Systems Laboratory, sponsored by the National Science Foundation, for the simulations and data analysis performed in this work. We also acknowledge the open-source Python community, and particularly the authors and contributors to the Matplotlib \citep{matplotlib}, NumPy \citep{numpy}, and MetPy \citep{metpy} packages that were used to generate many of the analyses and figures. Li and Chavas were supported by NSF grant AGS1648681. Reed was supported by NSF grant AGS1648629. 

%






%
%
%
\newcommand{\noopsort}[1]{} \newcommand{\printfirst}[2]{#1}
  \newcommand{\singleletter}[1]{#1} \newcommand{\switchargs}[2]{#2#1}

%

\end{document}